\documentclass[twocolumn,english,prx,superscriptaddress]{revtex4-2}
\usepackage{color}
\usepackage[utf8]{inputenc}
\usepackage{graphicx}
\usepackage[T1]{fontenc}
\usepackage{float}
\usepackage{natbib}
\usepackage{textcomp}
\usepackage{amsmath}
\usepackage{amssymb}
\usepackage{graphicx}
\usepackage{esint}
\usepackage{natbib}
\usepackage{xcolor}
\usepackage{multirow}

\newcommand{\kvec}{\vec{k}}

\usepackage[colorlinks=true,hyperfootnotes=true,breaklinks=true]{hyperref}
\usepackage[capitalise]{cleveref}

\begin{document}

\title{Superconducting gap symmetry of 2DEG at (111)-oriented LaAlO\textsubscript{3}/SrTiO\textsubscript{3} interface} 

\author{Julian Czarnecki}
\email{jczarnecki@agh.edu.pl}
\affiliation{AGH University of Krakow, Faculty of Physics and Applied Computer Science, al. A. Mickiewicza 30, 30-059 Krakow, Poland}

\author{Michał Zegrodnik}
\affiliation{AGH University of Krakow, Academic Centre for Materials and Nanotechnology, Al. Mickiewicza 30, 30-059 Krakow, Poland}


\author{Paweł W\'ojcik}
\email{pawel.wojcik@fis.agh.edu.pl}
\affiliation{AGH University of Krakow, Faculty of Physics and Applied Computer Science, al. A. Mickiewicza 30, 30-059 Krakow, Poland}

\begin{abstract}
We investigate the superconducting properties of the two-dimensional electron gas at the (111) LaAlO$_3$/SrTiO$_3$ interface. Using a multiorbital tight-binding model defined on a hexagonal lattice, we analyze the emergence of superconductivity driven by both interlayer (nearest-neighbor) and intralayer (next-nearest-neighbor) pairing interactions, with a particular focus on the symmetry of the superconducting gap. We demonstrate that, in both pairing scenarios, the superconducting gap transforms according to the $A_1$ irreducible representation of the $C_{6v}$ point group. Within the interlayer pairing scenario, the superconducting phase is characterized by a fully gapped quasiparticle excitation spectrum exhibiting extended $s$-wave symmetry, accompanied by an enhancement of the superconducting gap magnitude in the vicinity of the van Hove singularity. Conversely, the intralayer pairing channel produces a distinctive double-dome structure in the superconducting phase diagram, with the gap symmetry evolving from a fully gapped, extended $s$-wave at low carrier densities to a nodal extended $s$-wave state at higher electron concentrations. The qualitative agreement with experimentally observed nonmonotonic behavior of the critical temperature $T_c(V_g)$ suggests that intralayer next-nearest-neighbor pairing may play a dominant role in the superconductivity of the (111) LAO/STO interface.
\end{abstract}
\date{\today}
\maketitle

\section{Introduction}
Since the discovery of superconductivity at the LaAlO\textsubscript{3}/SrTiO\textsubscript{3} (LAO/STO) interface \cite{Ohtomo_Hwang_2004}, transition metal oxide heterostructures have attracted significant interest as a platform to study the interplay between 2D superconductivity~\cite{Reyren2007, joshua2012universal,maniv2015strong,
Biscaras, Monteiro2019, Monteiro2017}, spin-orbit coupling~\cite{Diez2015,Rout2017,Caviglia2010,Shalom2010,Yin2020,Singh2017,Hurand2015} and magnetism~\cite{Karen2012,Li2011,Brinkman2007,Dikin2011,Bert2011}. 
The extraordinary properties of such systems include not only an unusual phase diagram exhibiting a dome-shaped dependence of the critical temperature on the carrier density \cite{Caviglia_2008, Gariglio2015, Singh2024}, but also an anisotropy of the in-plane critical field, with magnitudes significantly exceeding the Chandrasekhar–Clogston limit \cite{Rout2017}. The observed anisotropy in (111) KTaO$_3$-based interfaces~\cite{Zhang2023} has been recently proposed to originate from $p$-wave pairing symmetry, sparking active debate about the symmetry of the superconducting gap in (111)-oriented transition metal oxide interfaces and their potential as platforms for realizing topological superconductivity~\cite{barthelemy2021, Fukaya2018} and Majorana bound states engineering~\cite{Lutchyn2018}.

Note, however, that for the widely studied LAO/STO interface, most experimental \cite{Caviglia_2008, Gariglio2015, Singh2024} and theoretical \cite{Zegrodnik_2020, Zegrodnik_2022, Boudjada_domes_2020, Lepori_interplay_singlet_triplet_2021, Yanase_multiorbital_2013} studies have focused on interfaces grown along the (001) crystallographic direction. In this configuration, the interface is described by a simple square lattice, and superconductivity is primarily attributed to the $d_{yz/xz}$ bands, with a negligible contribution from the lower-lying $d_{xy}$ orbitals~\cite{wojcik2021impact}. Our recent theoretical work has demonstrated that, for the (001) LAO/STO interface, the gap may exhibit an extended $s$-wave symmetry \cite{Wojcik_dome_schrodinger_poisson_sc_2024}, enabling semi-quantitative agreement with experimental observations~\cite{jouan_multiband_sc_2022} and successfully reproducing the dome-shaped dependence of the critical temperature on carrier concentration.

The crystallographic orientation of the LAO/STO interface, however, plays a pivotal role in determining the symmetry and hierarchy of its electronic structure. In the (111)-oriented configuration, the projection of the cubic perovskite lattice onto the (111) plane results in a quasi-two-dimensional system with an effective hexagonal symmetry~\cite{Xiao_interface_engineering_2011, Trama_2022, trama2023effect}. This reconstructed geometry gives rise to nontrivial modifications of the band structure and orbital hybridization. Recently, experimental evidence from angle-resolved photoemission spectroscopy (ARPES)~\cite{Rodel2014, Walker2014} and anisotropic magnetoresistance measurements~\cite{Rout2017_2} has confirmed the presence of sixfold rotational symmetry in the electronic spectrum. Interestingly, ARPES measurements reveal that all three $t_{2g}$ orbitals contribute comparably to the Fermi surface near the Brillouin zone center. This naturally eliminates the pivotal factor related to the Lifshitz transition and the occupation of a particular $d$ orbital, which is considered to be the origin of many physical phenomena occurring in the (001)-oriented LAO/STO, including the dome-like shape of $T_c$ as a function of carrier concentration~\cite{Trevisan2018}. Note, however, that in the (111)-oriented LAO/STO, a nonmonotonic, dome-shaped dependence of $T_c$ on gate voltage has also been reported~\cite{Rout2017,Monteiro2017,Monteiro2019,simon2024normal}. This property, together with others—such as the coexistence of two-dimensional superconductivity and magnetic order~\cite{Davis_magnetoresistance_2017}—still awaits a theoretical explanation within the framework of a realistic electronic structure and hexagonal crystalline symmetry, where the spectrum of possible induced gap symmetries is significantly richer~\cite{Pangburn_graphene_symmetries_2023}.

In this paper, we study the symmetry of the superconducting gap in a (111)-oriented LAO/STO interface, considering both interlayer and intralayer pairing mechanisms, within a real-space pairing scenario. Our self-consistent calculations, based on the realistic electronic structure modeled within the tight-binding approximation (TBA), have shown that the superconducting gap transforms according to the $A_1$ irreducible representation of the $C_{6v}$ point group. We found that
within the interlayer pairing scenario, the superconducting phase is characterized by a fully gapped quasiparticle excitation spectrum exhibiting extended $s$-wave symmetry. On the other hand, the intralayer pairing channel produces a distinctive double-dome structure in the superconducting phase diagram, with the gap symmetry evolving from a fully gapped extended $s$-wave at low carrier densities to nodal extended $s$-wave at higher electron concentrations. A qualitative comparison of our results with experimental data indicates that intralayer pairing between next-nearest neighbors may play a crucial role in superconductivity, contributing to the observed nonmonotonic dependence of $T_c(V_g)$.

The manuscript is organized as follows: in Sec. \ref{sec:Theoretical model} we introduce a theoretical model used to calculate superconducting properties of the (111) LAO/STO interface, and define the basic symmetry of the gap in this system, in Sec. \ref{sec:Results} we present results for intra-layer and inter-layer superconducting coupling scenarios, followed by a discussion. A summary is presented in Sec. \ref{sec:Summary}.

\section{Theoretical model}
\label{sec:Theoretical model}
\subsection{Tight-binding model}
\label{sec:Tight-binding model}
\begin{figure}[!ht]
    \includegraphics[width=0.5\columnwidth]{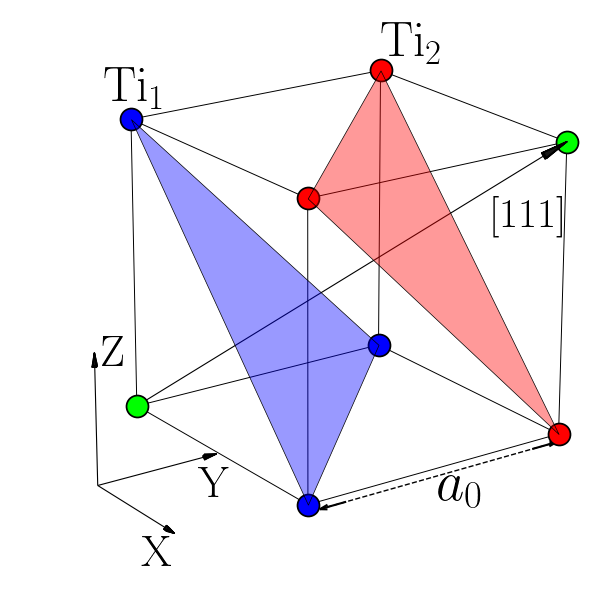}  \put(-120,110){(a)} \\
    \includegraphics[width=0.7\columnwidth]{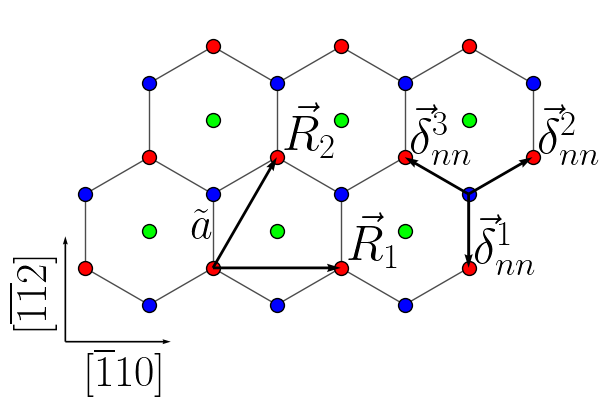}  \put(-140,110){(b)} 
\caption{(a) 3D structure of Ti atoms in the (111)-oriented LAO/STO interface together with (b) the projection of Ti atoms onto the (111) interface plane. Red, blue and green sites represent atoms belonging to different Ti layers, but only red and blue are considered in the bilayer scenario. Lattice constant  is denoted by $\tilde{a} = \sqrt{2 / 3}a_0$, where $a_0 = 0.39$~nm is the size of the 3D unit cell.}
\label{fig:hexagonal_111}
\end{figure}

To describe the LAO/STO interface in the (111) crystallographic direction, we project two layers of Ti atoms onto the (111) interface plane. The resulting hexagonal structure is shown in Fig.~~\ref{fig:hexagonal_111}, where the lattice vectors are defined as $\vec{R}_1 = \begin{pmatrix} \sqrt{3}, & 0\end{pmatrix}$ and $\vec{R}_2 = \begin{pmatrix} \frac{\sqrt{3}}{2}, & \frac{3}{2}\end{pmatrix}$ in units of lattice constant $\tilde{a}$.

\noindent Hamiltonian of the system is given by
\begin{equation}
    \hat{H} = \hat{H}_\text{TBA} + \hat{H}_\text{U} + \hat{H}_\text{SC},
\label{eq:H_general}
\end{equation}
where
\begin{equation}
    \hat{H}_\text{TBA} = \hat{H}_\text{0} +  \hat{H}_\text{SOC} + \hat{H}_\text{TRI} + \hat{H}_\text{v}
\label{eq:H_TBA}
\end{equation}
is a tight-binding Hamiltonian accounting for the electronic structure of 2DEG at the (111) LAO/STO interface~\cite{Trama_2022, trama2023effect, Trama_strain_2021}. In Eq.~(\ref{eq:H_TBA}), the kinetic term $\hat{H}_0$ takes the form
\begin{equation}
    \hat{H}_0 =  \sum_{\kvec} \sum_{l \alpha \sigma} \left ( t_{l}^{\alpha \overline{\alpha}} (\kvec) \hat{c}_{\kvec l \alpha \sigma }^\dag \hat{c}_{\kvec l \overline{\alpha} \sigma } - \mu \hat{c}_{\kvec l \alpha \sigma }^\dag \hat{c}_{\kvec l \alpha \sigma } \right ),
\label{eq:H_0}
\end{equation}
where $\mu$ is the chemical potential and the operators $\hat{c}^\dag_{\kvec l \alpha \sigma}$ ($\hat{c}_{\kvec l \alpha \sigma}$) create (anihilate) an electron in the state defined by the quantum numbers $\{\kvec l \alpha \sigma\}$, where $\kvec$ is the in-plane wave vector ($k_x := k_{\overline{1}10}$, $k_y := k_{\overline{11}2}$) belonging to the first Brillouin zone, $l = \{yz, zx, xy\}$ denotes the $d$ orbitals of Ti atoms, $\sigma = \{\uparrow, \downarrow\}$ is the electron spin and $\alpha = \{Ti_1, Ti_2\}$ is the layer index associated with the two Ti layers. In the considered model, the hopping terms are defined between two opposite Ti layers (nearest-neighbor hoppings), and are given by
{\small
\begin{subequations}
\begin{align}
    t_{yz}^{\alpha \overline{\alpha}}(\vec{k}) &= -t_\sigma  \left ( e^{i k_y} + e^{i \left( \frac{\sqrt{3}}{2}k_x - \frac{1}{2}k_y\right)}\right) - t_\pi e^{-i \left( \frac{\sqrt{3}}{2}k_x + \frac{1}{2}k_y \right)}, \\
    t_{zx}^{\alpha \overline{\alpha}}(\vec{k}) &= -t_\sigma \left ( e^{i k_y} + e^{-i \left( \frac{\sqrt{3}}{2}k_x + \frac{1}{2}k_y\right)}\right) - t_\pi e^{i \left( \frac{\sqrt{3}}{2}k_x - \frac{1}{2}k_y \right)}, \\
    t_{xy}^{\alpha \overline{\alpha}}(\vec{k}) &= -2 t_\sigma \cos \left( \frac{\sqrt{3}}{2} k_x \right) e^{-i\frac{1}{2} k_y} - t_\pi e^{i k_y},
\label{eq:t_hoppings}
\end{align}
\end{subequations}
}
where $t_\sigma$ and $t_\pi$ correspond to $\sigma$ and $\pi$ type overlap between the $d_{xy,yz,xz}$ atomic orbitals.\\

The atomic spin-orbit (SO) coupling at the (111) LAO/STO interface is taken into account by the term
\begin{equation}
    \hat{H}_\text{SOC} = \frac{\lambda}{2} \sum_{\kvec} \sum_{l,l',m, \alpha, \sigma, \sigma'} i \epsilon_{l, l', m} \hat{c}_{\kvec l \alpha \sigma}^\dag \sigma_{\sigma \sigma'}^{m} \hat{c}_{\kvec l' \alpha \sigma'},
\label{eq:H_SOC}
\end{equation}
where $\epsilon$ is a Levi--Civita tensor, $\sigma^m$ is one of the Pauli matrices $\sigma^m = \{ \sigma^x, \sigma^y, \sigma^z\}$ and $\lambda$ determines the strength of the SO coupling. Additionally, Hamiltonian (\ref{eq:H_TBA}) includes the effect of trigonal strain at the interface
\begin{equation}
    \hat{H}_\text{TRI} = \frac{\Delta_{TRI}}{2} \sum_{\kvec} \sum_{l,l' \neq l, \alpha, \sigma} \hat{c}_{\kvec l \alpha \sigma}^\dag \hat{c}_{\kvec l' \alpha \sigma},
\end{equation}
where the sign of $\Delta_\text{TRI}$ determines whether the distance between layers is contracted ($\Delta_{TRI} < 0$) or dilated ($\Delta_{TRI} > 0$). Finally, the breaking of inversion symmetry between the two Ti layers, which induces an electric field perpendicular to the plane, is incorporated through the term 
\begin{equation}
\begin{aligned}
    \hat{H}_\text{v}=&\frac{v}{2} \sum_{\kvec} \sum_{l \alpha \sigma} \xi_\alpha \hat{c}_{\kvec l \alpha \sigma}^\dag \hat{c}_{\kvec l \alpha \sigma} + \\
    &\sum_{\kvec} \sum_{ll' \alpha \alpha' \sigma} h_{\kvec l l'}^{\alpha \alpha'}\left( v \right)\hat{c}_{\kvec l \alpha \sigma}^\dag \hat{c}_{\kvec l' \alpha' \sigma},
\label{eq:H_externalElectric}    
\end{aligned}
\end{equation}
where $v$ is a parameter describing the magnitude of the electric field, $\xi_{Ti_1} = +1$ and $\xi_{Ti_2} = -1$ account for the potential offset between layers and  $h_{\kvec l l'}^{\alpha \alpha'} \left ( v \right )$ leads to renormalization of the hopping terms in the presence of the electric field. For details regarding the form of the latter, see Appendix~\ref{sec:HoppingRenormalization}.\\

In our model, the Coulomb repulsion between electrons is included in the form of the Hubbard term, which can be written as
\begin{equation}
    \hat{H}_\text{U} = U \sum_{i l \sigma} \hat{n}_{il\sigma} \hat{n}_{il \overline{\sigma}} + V\sum_{i l' \neq l} \hat{n}_{il}\hat{n}_{il'},
\label{eq:Hubbard_initial}
\end{equation}
where the energies $U$ and $V$ denote intra- and interorbital Coulomb integrals and $\hat{n}_{il} = \hat{n}_{il \uparrow} + \hat{n}_{il \downarrow}$ is the particle number operator on the site $i$ in the orbital $l$. By applying the standard mean-field approximation and transforming to reciprocal space we obtain
\begin{equation}
    \hat{H}_\text{U} = \sum_{\kvec} \sum_{l \alpha \sigma} \hat{n}_{\kvec l \alpha \sigma } \vphantom{\sum_{l' \neq l}} 
    \left[ U n_{l \alpha \overline{\sigma} } + V \sum_{l' \neq l} \left(   n_{l' \alpha \sigma }+  
     n_{l' \alpha \overline{\sigma}} \right) \right] + \gamma_U,
\end{equation}
where
\begin{equation}
     n_{l \alpha \sigma}  \equiv \sum_{\kvec} \langle \hat{c}_{\kvec l \alpha \sigma}^\dag \hat{c}_{\kvec l \alpha \sigma} \rangle
\label{eq:N_expected}
\end{equation}
is an average number of particles per site, and 
\begin{equation}
    \gamma_\text{U} = -\sum_{\kvec l \alpha \sigma}  n_{\kvec l \alpha \sigma} \left [ U  n_{\kvec l \alpha \overline{\sigma}} + \frac{1}{2}V \sum_{l' \neq l} \left(  n_{\kvec l \alpha \sigma} + n_{\kvec l \alpha \overline{\sigma}} \right) \right].
\label{eq:gammaU}
\end{equation}
\\

We assume that the superconducting state is induced
by a real-space intraorbital pairing between the nearest neighboring and next-nearest neighboring sites, as well as the interorbital Cooper pair hopping 
\begin{equation}
\begin{aligned}
    \hat{H}_\text{SC} =& -\sum_{i, j > i, l, \sigma} J_{ij} \hat{c}_{il\sigma}^\dag \hat{c}_{jl\bar{\sigma}}^\dag \hat{c}_{il\bar{\sigma}} \hat{c}_{jl\sigma}  \\
    &- \sum_{i, j > i, \sigma} J'_{ij} \sum_{l' \neq l} \hat{c}_{il\sigma}^\dag \hat{c}_{jl\bar{\sigma}}^\dag \hat{c}_{il'\bar{\sigma}} \hat{c}_{jl'\sigma},
\label{eq:H_SC}
\end{aligned}
\end{equation}
where $J_{ij}$ denotes intraorbital superconducting coupling strength between electrons occupying sites $i$ and $j$, while $J'_{ij}$ is the interorbital Cooper pair hopping strength between sites $i$ and $j$.
Note that the projection of a bilayer of Ti atoms onto (111) plane results in a hexagonal lattice (see Fig.~\ref{fig:hexagonal_111}), where the nearest neighbors (NN) belong to different Ti layers, while next-nearest neighbors (NNN) reside within the same Ti layer. In the further part of the paper we use the terms intra- and interlayer coupling to distinguish between NN and NNN coupling and denote their strength by $J$ and $J_{nnn}$, respectively. \\

By applying the standard mean-field approximation, the Hamiltonian $H_\text{SC}$ in the reciprocal space can be written in the form
\begin{eqnarray}
    \hat{H}_\text{SC} = \sum_{\kvec l \alpha \sigma} && \big [ \Gamma_{l \alpha \overline{\alpha}}^{\sigma \overline{\sigma}} \left( \kvec \right) \hat{c}_{\kvec l \alpha \sigma}^\dag \hat{c}_{-\kvec l \bar{\alpha} \bar{\sigma}}^\dag \label{eq:H_SC_final} \\
    &+& \Gamma_{l \alpha \alpha}^{\sigma \overline{\sigma}} \left( \kvec \right) \hat{c}_{\kvec l \alpha \sigma}^\dag \hat{c}_{-\kvec l \alpha \bar{\sigma}}^\dag \big ] + h.c. + \gamma_\text{SC}, \nonumber
\end{eqnarray}    
where
\begin{subequations}
\begin{align}
    \Gamma_{l \alpha \overline{\alpha}}^{\sigma \overline{\sigma}} \left( \kvec \right) &= \sum_{\vec{\delta}_{nn}} \Gamma_{l \alpha \overline{\alpha}}^{\sigma \overline{\sigma}}(\vec{\delta}_{nn}) e^{-i \kvec \vec{\delta}_{nn}} 
    = \sum_{\vec{\delta}_{nn}} e^{-i \kvec \vec{\delta}_{nn}} \times  \label{eq:gamma_nn}  \\
    &\left [ -\frac{J}{2}\Delta_{l \alpha \bar{\alpha}}^{\bar{\sigma} \sigma}\left ( \vec{\delta}_{nn} \right)  - \frac{J'}{2} \sum_{l' \neq l} \Delta_{l' \alpha \bar{\alpha}}^{\bar{\sigma} \sigma}(\vec{\delta}_{nn}) \right] , \nonumber \\
    \Gamma_{l \alpha \alpha}^{\sigma \overline{\sigma}} \left( \kvec \right) &= \sum_{\vec{\delta}_{nnn}} \Gamma_{l \alpha \alpha}^{\sigma \overline{\sigma}}(\vec{\delta}_{nnn}) e^{-i \kvec \vec{\delta}_{nnn}} 
    = \sum_{\vec{\delta}_{nnn}}  e^{-i \kvec \vec{\delta}_{nnn}} \times \label{eq:gamma_nnn} \\
    &\left[ -\frac{J_{nnn}}{2}\Delta_{l \alpha \alpha}^{\bar{\sigma} \sigma}\left ( \vec{\delta}_{nnn} \right) \right. \nonumber - \left. \frac{J_{nnn}'}{2} \sum_{l' \neq l} \Delta_{l' \alpha \alpha}^{\bar{\sigma} \sigma}(\vec{\delta}_{nnn}) \right]. \nonumber
\end{align}
\end{subequations}    
The last term in (\ref{eq:H_SC_final}) represents the reference energy and has the form 
\begin{equation}
\begin{aligned}
    \gamma_\text{SC} := & \frac{1}{2} \sum_{\vec{\delta}_{nn}}\sum_{ l \alpha \sigma} \left ( \Delta_{l \alpha \bar{\alpha}}^{\sigma \bar{\sigma}} (\vec{\delta}_{nn}) \right )^{\dag}  \times \\
    & \left [ J \Delta_{l \alpha \bar{\alpha}}^{\sigma \bar{\sigma}} (\vec{\delta}_{nn})  + J' \sum_{l' \neq l}  \Delta_{l' \alpha \bar{\alpha}}^{\sigma \bar{\sigma}} (\vec{\delta}_{nn})\right ] + \\
    & \frac{1}{2} \sum_{\vec{\delta}_{nnn}} \sum_{ l \alpha \sigma} \left ( \Delta_{l \alpha \alpha}^{\sigma \bar{\sigma}} (\vec{\delta}_{nnn}) \right )^{\dag}  \times \\
    &\left [ J_{nnn} \Delta_{l \alpha \alpha}^{\sigma \bar{\sigma}} (\vec{\delta}_{nnn})  + J'_{nnn} \sum_{l' \neq l}  \Delta_{l' \alpha \alpha}^{\sigma \bar{\sigma}} (\vec{\delta}_{nnn})\right ].
\label{eq:gamma_SC}
\end{aligned}
\end{equation}
In the above equations, we have defined the pairing amplitudes  
\begin{equation}
\begin{aligned}
    \Delta_{l \alpha_i \alpha_j}^{\sigma \bar{\sigma}} (\vec{\delta}_{ij})  := \frac{1}{N} \sum_{ij \kvec'} \langle \hat{c}_{\kvec' l\sigma \alpha_i}^\dag \hat{c}_{\kvec' l \bar{\sigma} \alpha_j}^\dag \rangle \cdot e^{-i \kvec' \left( \vec{r}_i - \vec{r}_j \right)},
\label{eq:mean_pairing}
\end{aligned}
\end{equation}
where $\vec{\delta}_{ij}=\vec{r}_i - \vec{r}_j$. A vector connecting any site with its nearest (next-nearest) neighbors is denoted by $\vec{\delta}_{nn} (\vec{\delta}_{nnn})$ and belongs to the set
\begin{subequations}
\begin{align}
    \vec{\delta}_{nn} &= \Bigl\{ \pm \begin{pmatrix} 0 & -1 \end{pmatrix}^T , \pm \begin{pmatrix} \frac{\sqrt{3}}{2} & \frac{1}{2} \end{pmatrix}^T, \pm \begin{pmatrix} -\frac{\sqrt{3}}{2} & \frac{1}{2} \end{pmatrix}^T \Bigr\} \label{eq:delta_nn} \\
    &= \Bigl\{ \pm \vec{\delta}_{nn}^{(1)} , \pm \vec{\delta}_{nn}^{(2)}, \pm \vec{\delta}_{nn}^{(3)} \Bigr\}, \nonumber \\
    \vec{\delta}_{nnn} &= \Bigl\{ \pm \begin{pmatrix} \sqrt{3} & 0 \end{pmatrix}^T ,  \pm \begin{pmatrix} \frac{\sqrt{3}}{2} & \frac{3}{2} \end{pmatrix}^T, \pm \begin{pmatrix} -\frac{\sqrt{3}}{2} & \frac{3}{2} \end{pmatrix}^T \Bigr\} \label{eq:delta_nnn} \\
    &= \Bigl\{ \pm \vec{\delta}_{nnn}^{(1)} , \pm \vec{\delta}_{nnn}^{(2)}, \pm \vec{\delta}_{nnn}^{(3)} \Bigr\}. \nonumber
    \end{align}
\end{subequations}
One should note that the change of sign of any $\vec{\delta}_{nn}$ leads to the inversion of lattice indices $\alpha$ in Eq. \eqref{eq:H_SC_final}.

Finally, by transforming the Hamiltonian ~\eqref{eq:H_general} into Nambu space, we obtain
\begin{eqnarray}
    \hat{H} &=& \frac{1}{2} \sum_{\kvec}  \vec{f}_{\kvec}^{~\dag} \mathcal{H} \left ( \kvec \right ) \vec{f}_{\kvec} + \gamma_\text{SC} + \gamma_\text{U} \nonumber \\
    &+& \frac{1}{2} \sum_{\kvec} Tr\left( \hat{H}_\text{TBA}\left ( -\kvec \right ) + \hat{H}_\text{U}\left(-\kvec \right)\right)  ,
\label{eq:H_Nambu}
\end{eqnarray}
where $Tr(...)$ is a trace operator, the vector $\vec{f}_{\kvec}= \left( \vec{c}_{\kvec l \alpha \sigma}, \vec{c}_{-\kvec l \alpha \overline{\sigma}}^{~\dag} \right)$,
and the Hamiltonian $\mathcal{H}(\kvec)$ has the form of  $24\times 24$ matrix 
\begingroup
\small
\begin{equation}
    \mathcal{H}(\kvec) = \begin{pmatrix}
        H_\text{TBA}(\kvec)  + H_\text{U}(\kvec) & \Gamma (\kvec)\\
        \Gamma^\dag (\kvec) & -\left( H_\text{TBA}(-\kvec)  + H_\text{U}(-\kvec) \right )^\dag
    \end{pmatrix}.
\label{eq:H_BdG}
\end{equation}
\endgroup
In Eq.~\eqref{eq:H_BdG}, $\Gamma (\kvec)$ is the $12\times 12$ matrix constructed from elements given by Eqs.~\eqref{eq:gamma_nn} and \eqref{eq:gamma_nnn}. Their values are determined in a self-consistent way, which requires the integration over the first Brillouin zone in the wave vector space - see Eqs.~\eqref{eq:N_expected} and \eqref{eq:mean_pairing}. We assume that self-consistency is reached when the difference in the superconducting matrix elements \eqref{eq:gamma_nn} \eqref{eq:gamma_nnn} between two consecutive iterations is less than $10^{-4}$~meV.\\

The calculations have been performed using the following material parameters corresponding to (111) LAO/STO interface~\cite{Trama_2022}: $t_{\sigma} = 500$~meV, $t_{\pi} = 40$~eV, $\lambda = 10$~meV, $\Delta_\text{TRI} = -5$~meV, $v = 200$~meV, $V_{pd\pi} = 28$~meV, $V_{pd\sigma} = -65$~meV, $U = V = 2$~eV. Initially, matrix elements $\Gamma$ are all set to the same magnitude in a pure spin singlet state $\Gamma^{\sigma \overline{\sigma}} = -\Gamma^{\overline{\sigma} \sigma} $. In the numerical procedure, to accelerate the integration, we use Romberg's numerical scheme, while the convergence is improved by utilizing the Broyden's method.

\begingroup
\begin{table}[htbp]
\caption{\label{tab:CharacterTable} Character table of point group $C_{6v}$.}
\begin{ruledtabular}
\begin{tabular}{ccccccc}
& E & 2C\textsubscript{6} & 2C\textsubscript{3} & C\textsubscript{2} & 3$\sigma_v$ & 3$\sigma_d$ \\
\hline
A\textsubscript{1} & 1 & 1 & 1 & 1 & 1 & 1 \\
A\textsubscript{2} & 1 & 1 & 1 & 1 & -1 & -1 \\
B\textsubscript{1} & 1 & -1 & 1 & -1 & 1 & -1 \\
B\textsubscript{2} & 1 & -1 & 1 & -1 & -1 & 1 \\
E\textsubscript{1} & 2 & 1 & -1 & -2 & 0 & 0 \\
E\textsubscript{2} & 2 & -1 & -1 & 2 & 0 & 0 \\
\end{tabular}    
\end{ruledtabular}

\end{table}
\endgroup  

\subsection{Symmetry analysis}
\label{sec:Symmetry analysis}
Based on the self-consistently determined pairing amplitudes in the real space $\Gamma_{l \alpha \alpha'}^{\sigma \overline{\sigma}}  ( \vec{\delta}  )$, we analyze the symmetry of the superconducting gap. Although the lattice presented in Fig. \ref{fig:hexagonal_111} has a $C_{6v}$ symmetry, we must note that the term (\ref{eq:H_externalElectric}) lowers the symmetry of the Hamiltonian to $C_{3v}$, by introducing a potential difference between the sublattices $Ti_1$ and $Ti_2$. The latter point group does not contain the in-plane inversion symmetry $C_2: \vec{r} \rightarrow -\vec{r}$, nor any other symmetry element that couples the two sublattices. This fact has two important implications: (i) parity is not imposed by the symmetry of the Hamiltonian, which in a general case can lead to mixing of singlet and triplet pairing channels \cite{Smidman_ASOC_review_2017}; (ii) symmetry decomposition in $C_{3v}$ point group gives no clear evidence whether a state is even or odd, which is of main importance to distinguish between singlet and triplet pairing states. To resolve the second issue, we classify the states with respect to the irreducible representations (IRs) of the $C_{6v}$ point group, in which the notion of parity is already embedded. This approach is justified by the fact that we restrict our analysis to purely singlet pairing, which leads to states with well-defined inversion symmetry.

For the sake of classification, we construct projection operators \cite{Senechal2019, Senechal2020, Chen2020} 
\begin{equation}
    P^{(\xi)} = \frac{1}{|G|}\sum_g d_\xi \chi^{(\xi)*}(g) R(g),
    \label{eq:ProjectionOperator}
\end{equation}
where $\xi$ denotes the IR, $|G|$ is the number of elements in $C_{6v}$ group, $d_{\xi}$ is the dimension of IR ($\xi$), $\chi^{(\xi)}(g)$ is the character of operation $g$ in IR and $R(g)$ is the (matrix) representation of the operation $g$. The character table, which contains all possible IRs and conjugacy classes, is given in Table~\ref{tab:CharacterTable}.

\begingroup
\begin{table}[htbp]
\caption{\label{tab:GeneratorsC6v} Matrix representations of generators of $C_{6v}$ point group acting on spatial ($\mathcal{R}$) and orbital ($\mathcal{O}$) submodules. $C_6$ represents a rotation by $\frac{\pi}{3}$ and $\sigma_{v_1}$ is a mirror with respect to the $y=0$ (diagonal of a hexagon).}
\begin{ruledtabular}
\begin{tabular}{ccc}
 & $R(C_6)$ & $R(\sigma_{v_1})$ \\
\hline \\
$\mathcal{O}$ & 
    $\begin{pmatrix}
    0 & 0 & 1 \\
    1 & 0 & 0 \\
    0 & 1 & 0
    \end{pmatrix}$ & 
    $\begin{pmatrix}
    0 & 1 & 0 \\
    1 & 0 & 0 \\
    0 & 0 & 1
    \end{pmatrix}$ \\ \\
$\mathcal{R}$ & 
    $\begin{pmatrix}
    0 & 1 & 0 & 0 & 0 & 0 \\
    0 & 0 & 1 & 0 & 0 & 0 \\
    0 & 0 & 0 & 1 & 0 & 0 \\
    0 & 0 & 0 & 0 & 1 & 0 \\
    0 & 0 & 0 & 0 & 0 & 1 \\
    1 & 0 & 0 & 0 & 0 & 0
    \end{pmatrix}$ & 
    $\begin{pmatrix}
    1 & 0 & 0 & 0 & 0 & 0 \\
    0 & 0 & 0 & 0 & 0 & 1 \\
    0 & 0 & 0 & 0 & 1 & 0 \\
    0 & 0 & 0 & 1 & 0 & 0 \\
    0 & 0 & 1 & 0 & 0 & 0 \\
    0 & 1 & 0 & 0 & 0 & 0
    \end{pmatrix}$ \\
\end{tabular}
  
\end{ruledtabular}

\end{table}
\endgroup
To construct appropriate representations $R(g)$ acting on a full module $\mathcal{T}$, we shall consider the action of the symmetry operations on all three submodules, related to the degrees of freedom of our system: $\mathcal{R}$ (spatial), $\mathcal{O}$ (orbital) and $\mathcal{S}$ (spin). Since the spin-orbit coupling in our system is assumed to be much smaller than the kinetic term  $\lambda \ll t_{\sigma}$, and only spin-singlet pairing is allowed in the coupling mechanism, we restrict our analysis to the spatial and orbital components of the order parameter. This leads to a product in the form 
\begin{equation}
    \mathcal{T} = \mathcal{O} \otimes \mathcal{R}.
\label{eq:fullModule}
\end{equation}
In module $\mathcal{T}$ we can write representations as
\begin{equation}
    R^{\mathcal{T}}(g) = R^{\mathcal{O}}(g) \otimes R^{\mathcal{R}}(g),
\end{equation}
where $R^{\mathcal{O}}(g)$ is the representation acting on the module $\mathcal{O}$ in basis $(d_{yz},d_{zx},d_{xy})^T$, while $R^{\mathcal{R}}(g)$ acts on the module $\mathcal{R}$ in basis $(\vec{\delta}_{nn}^{(1)}, -\vec{\delta}_{nn}^{(3)}, \vec{\delta}_{nn}^{(2)}, -\vec{\delta}_{nn}^{(1)}, \vec{\delta}_{nn}^{(3)}, -\vec{\delta}_{nn}^{(2)})^T$ for nearest neighbors or $(\vec{\delta}_{nnn}^{(1)} , \vec{\delta}_{nnn}^{(2)},  \vec{\delta}_{nnn}^{(3)} , -\vec{\delta}_{nnn}^{(1)},  -\vec{\delta}_{nnn}^{(2)},  -\vec{\delta}_{nnn}^{(3)} )^T$ for next-nearest neighbors. Generators of the $C_{6v}$ point group in the aforementioned bases are given in Table \ref{tab:GeneratorsC6v}. 

As a result, we construct the vector of the order parameter amplitudes according to our basis \eqref{eq:fullModule}
\begin{equation}
    | \psi^\Gamma \rangle = \begin{pmatrix}
        \Gamma^{\sigma \overline{\sigma}}_{d_{yz}} \left( \vec{\delta}^1 \right) & \Gamma^{\sigma \overline{\sigma}}_{d_{yz}} \left( \vec{\delta}^2 \right) & ... & \Gamma^{\sigma \overline{\sigma}}_{d_{zx}} \left( \vec{\delta}^1 \right) & ...
    \end{pmatrix}^T,
\end{equation}
where, for simplicity, we have omitted the $nn$ and $nnn$ subscripts. We calculate the eigenvalues and eigenvectors of projection operators \eqref{eq:ProjectionOperator} for each IR and choose the eigenvectors corresponding to the eigenvalue $1$, denoted as $| \psi^{\xi^{(n)}} \rangle$. Note that each IR might have several such eigenvectors, depending on its multiplicity in the full module, which is given by
\begin{equation}
    N_\xi = \frac{1}{|G|} \sum_g \chi^{(\xi)*}(g)Tr\left( R^\mathcal{T}(g) \right).
\end{equation}
Moreover, two IRs of $C_{6v}$ are two dimensional, namely $E_1$ and $E_2$, which further multiplies the number of distinct eigenvectors by IR's dimension $d_\xi$. As a consequence, apart from an IR label ($\xi$), we  add index $n$ that differentiates between eigenvectors that correspond to the same IR. Finally, we compute the projection
\begin{equation}
    \Gamma^{\sigma \overline{\sigma}\xi^{(n)}}_{\alpha \alpha'} = \frac{\left | \langle \psi^{\xi^{(n)}} | \psi^\Gamma \rangle \right |}{\dim \left(\psi^\Gamma\right)},
\label{eq:projectionEnergy}
\end{equation}
where $\dim \left(\psi^\Gamma\right)$ is the dimension of vector of order parameters. This provides information on the energy corresponding to a given IR's eigenvector ($\xi^{(n)}$) per orbital per bond. Since in our model we assume the spin-singlet pairing, $\Gamma^{\sigma \overline{\sigma}} = -\Gamma^{\overline{\sigma} \sigma}$,  $\Gamma^{\sigma \overline{\sigma}\xi^{(n)}}_{\alpha \alpha'}$ is invariant under spin inversion (due to the absolute value in Eq. \ref{eq:projectionEnergy}). For this reason, we omit the spin indices and henceforth denote the projected energy corresponding to a given IR by $\Gamma^{\xi^{(n)}}_{\alpha \alpha'}$. The latter defines two superconducting coupling channels: interlayer, related to the superconducting coupling of the nearest neighbors and denoted by $\Gamma^{\xi^{(n)}}_{\alpha \overline{\alpha}}$; and intralayer, related to the coupling to the next-nearest neighbors and denoted by $\Gamma^{\xi^{(n)}}_{\alpha \alpha}$.

\section{Results}
\label{sec:Results}
\begin{figure}[!t]
\includegraphics[width=0.5\columnwidth]{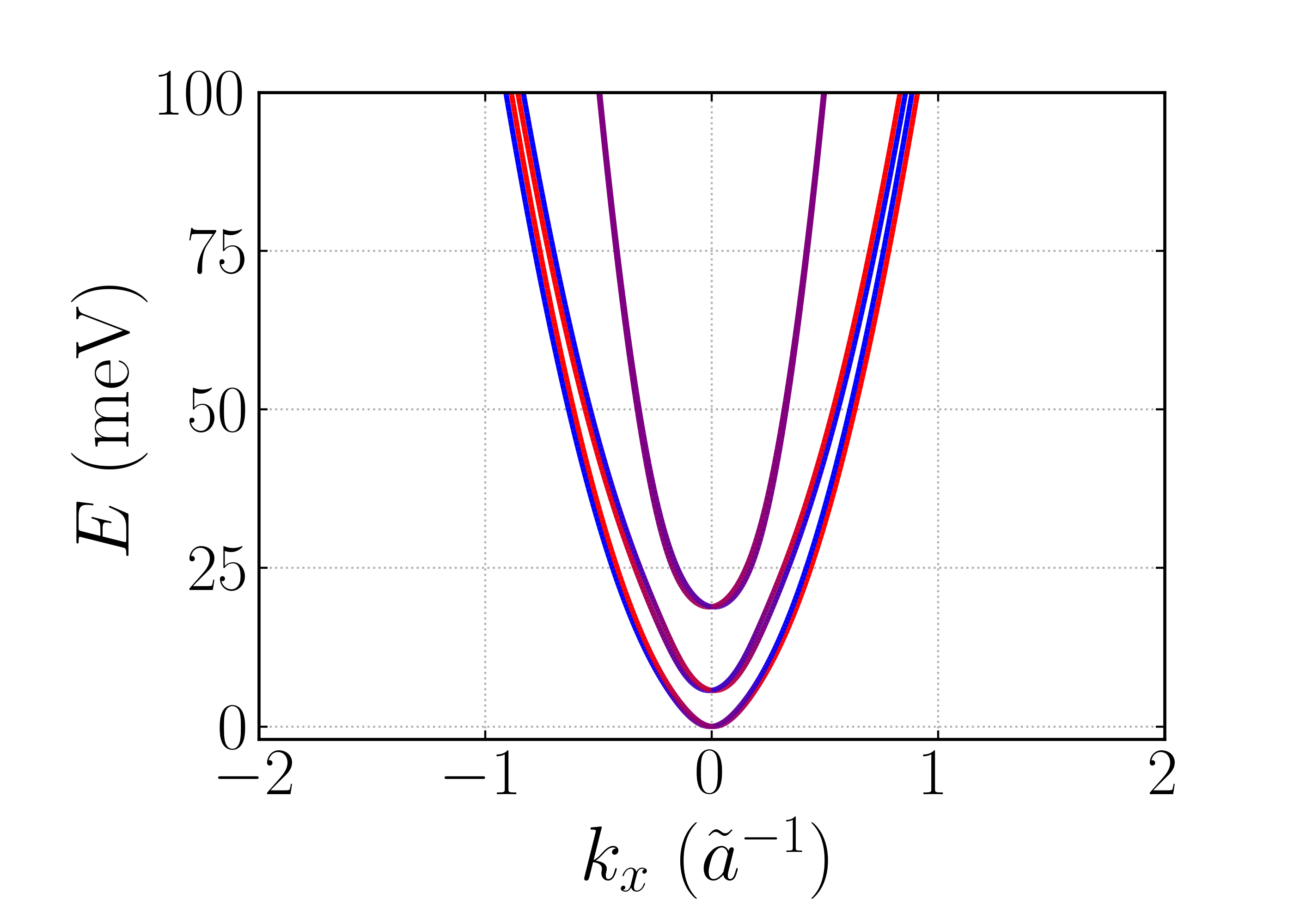}  \put(-95,69){(a)}
\includegraphics[width=0.5\columnwidth]{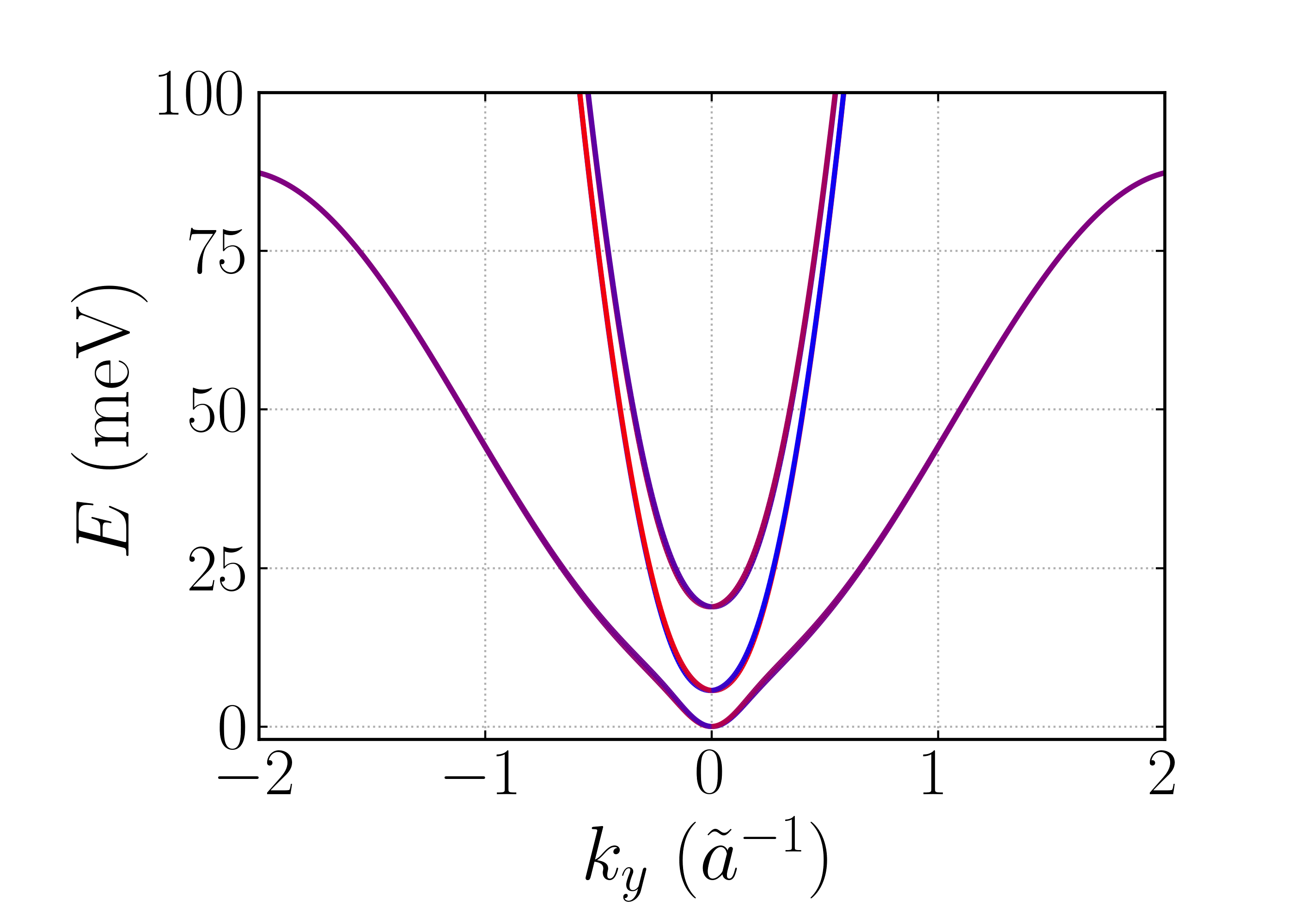}  \put(-95,69){(b)} \\
\includegraphics[width=0.5\columnwidth]{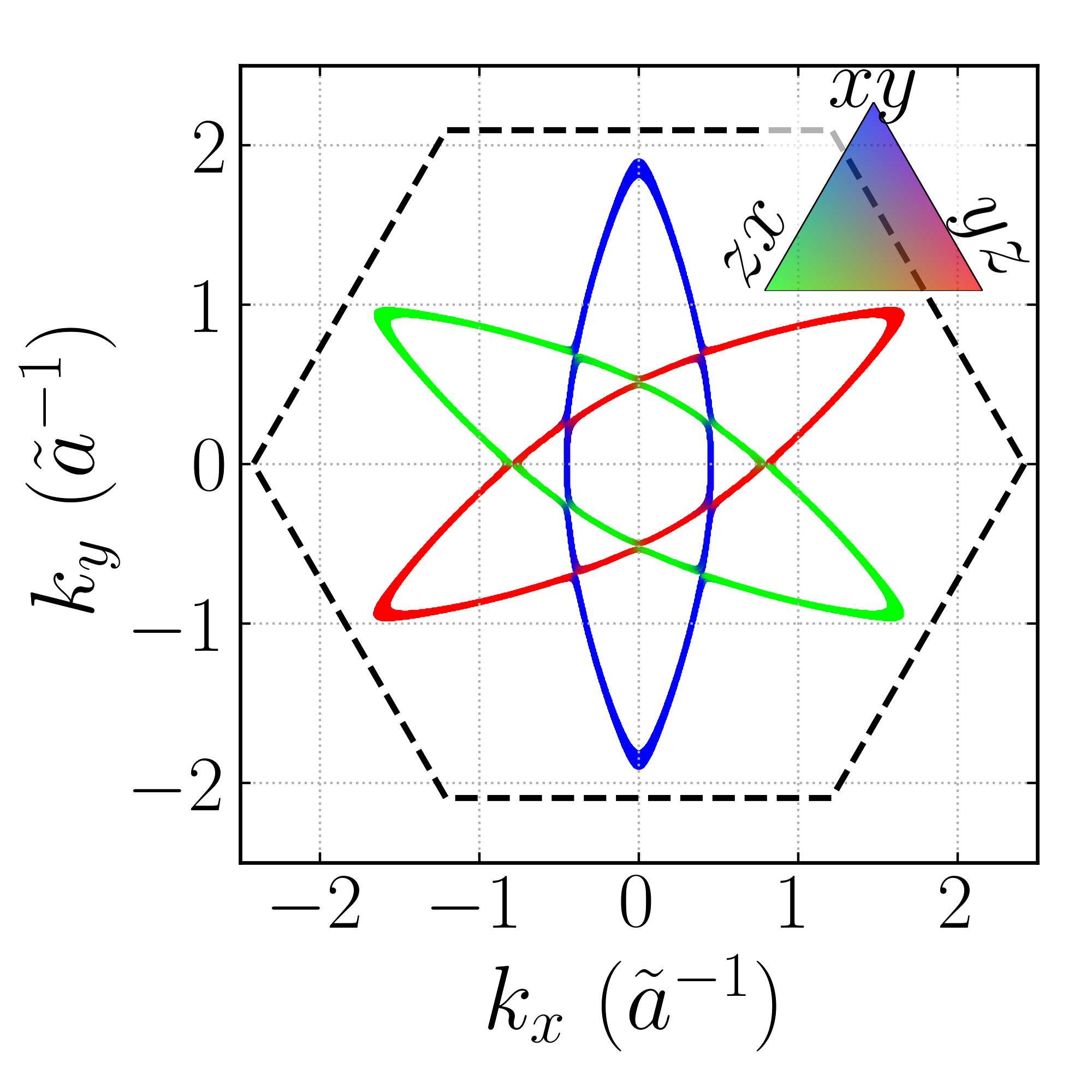}  \put(-92,105){(c)} 
\includegraphics[width=0.5\columnwidth]{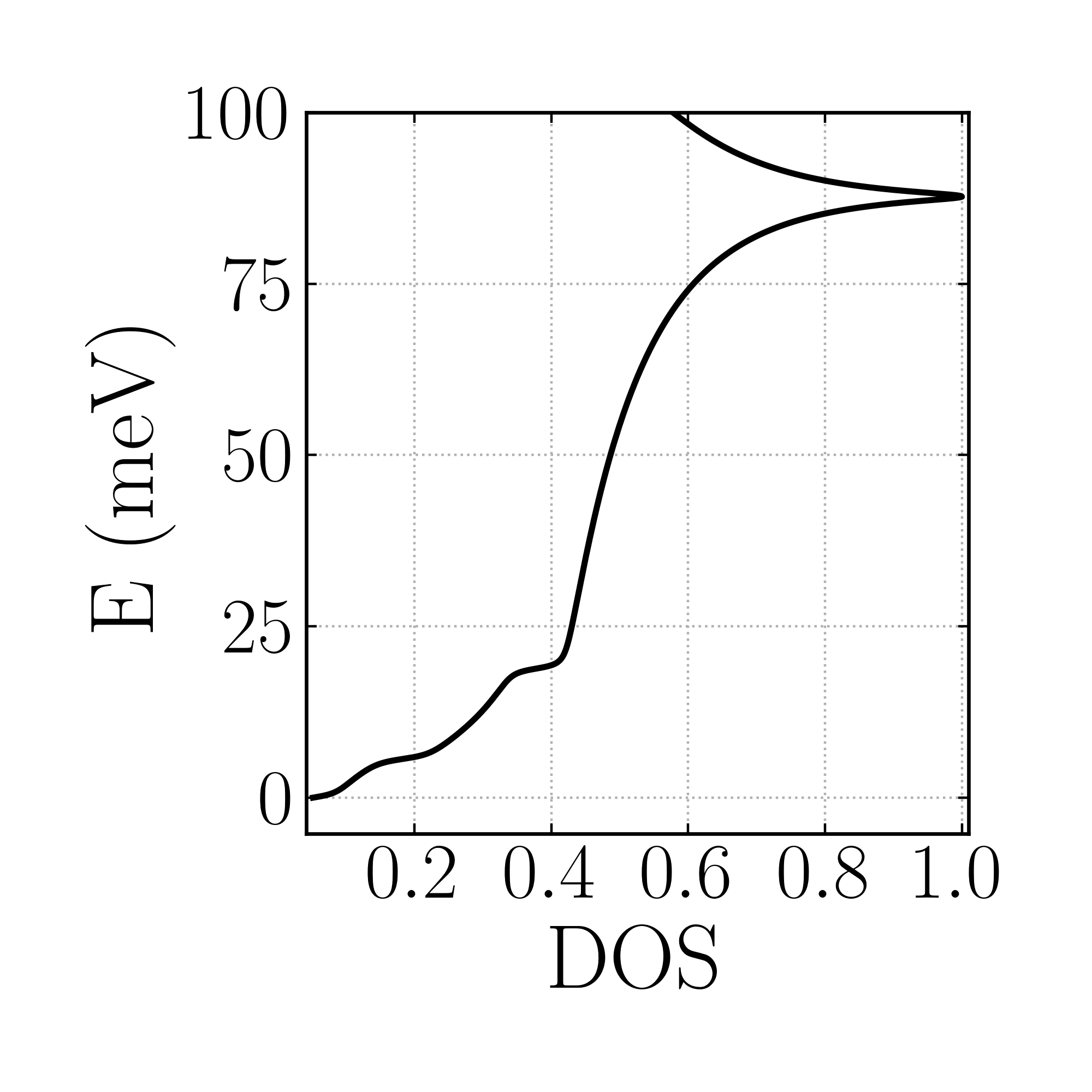}  \put(-85,100){(d)}
\caption{(a, b) Dispersion relations of 2DEG at the (111) LAO/STO interface determined along the $X$ and $Y$ directions of the Brillouin zone. The color of the lines indicates the spin $z$ component, $\uparrow$ - red, $\downarrow$ - blue. (c) Fermi surface for $\mu=85$~meV with contributions from $d$-type orbitals marked by color: red -- $yz$, green -- $zx$, blue -- $xy$. (d) Density of states (DOS) with visible van Hove singularity around $E=85$~meV.}
\label{fig:dispersions}
\end{figure}
\begin{figure*}[!t]
    \centering
    \includegraphics[width=0.33\linewidth]{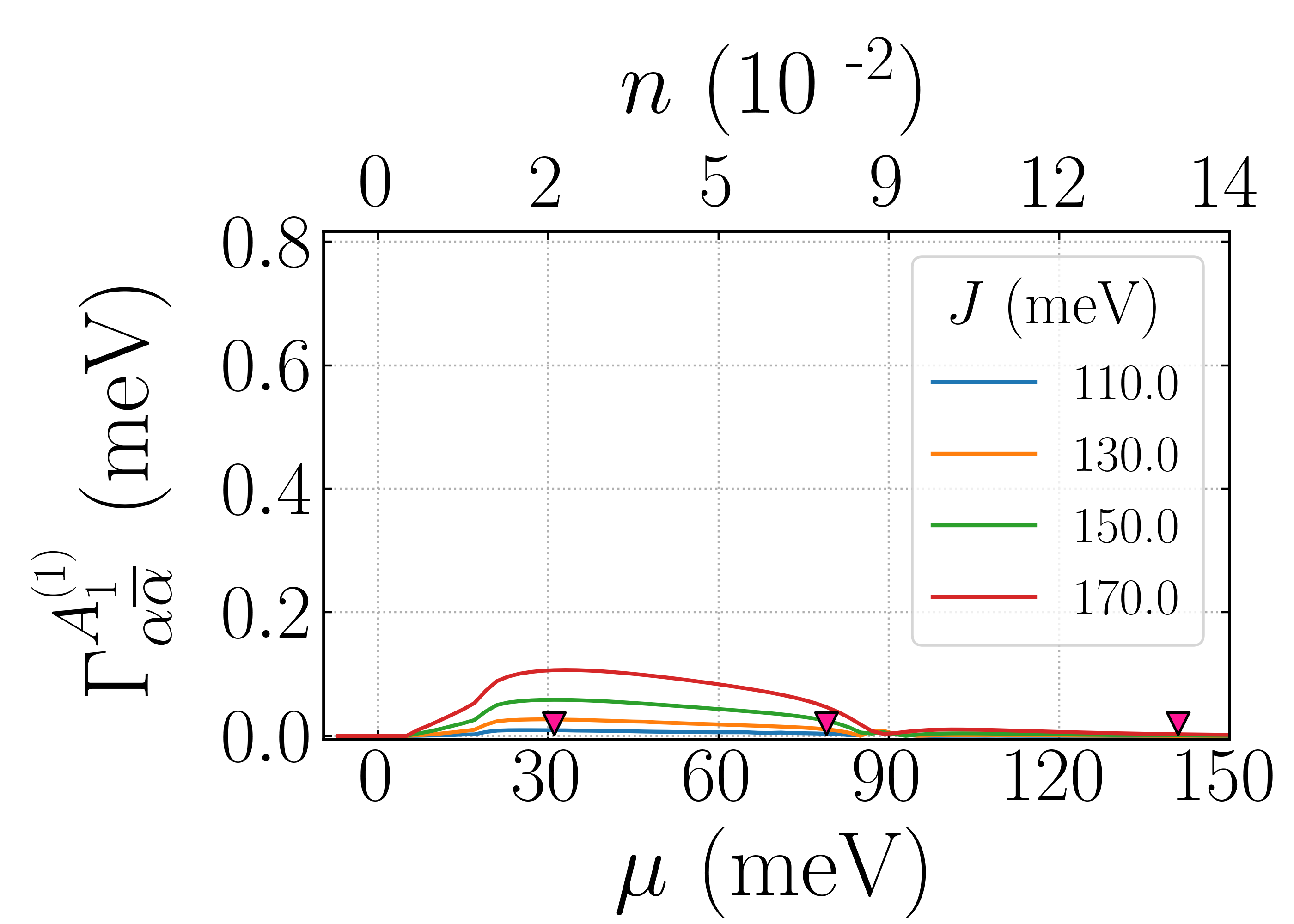}
    \put(-150,107){(a)}
    \includegraphics[width=0.33\linewidth]{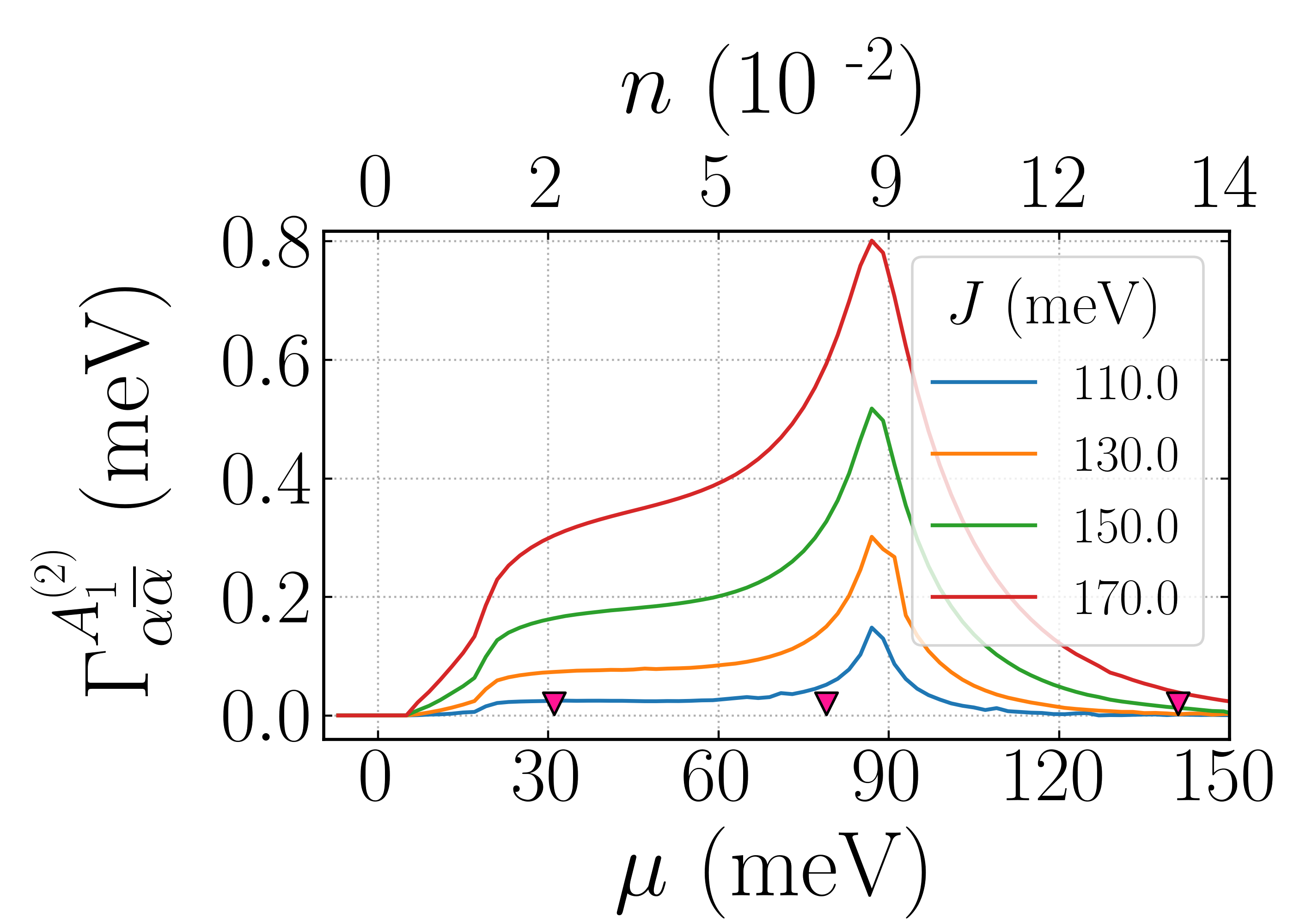}
    \put(-150,107){(b)}
    \includegraphics[width=0.33\linewidth]{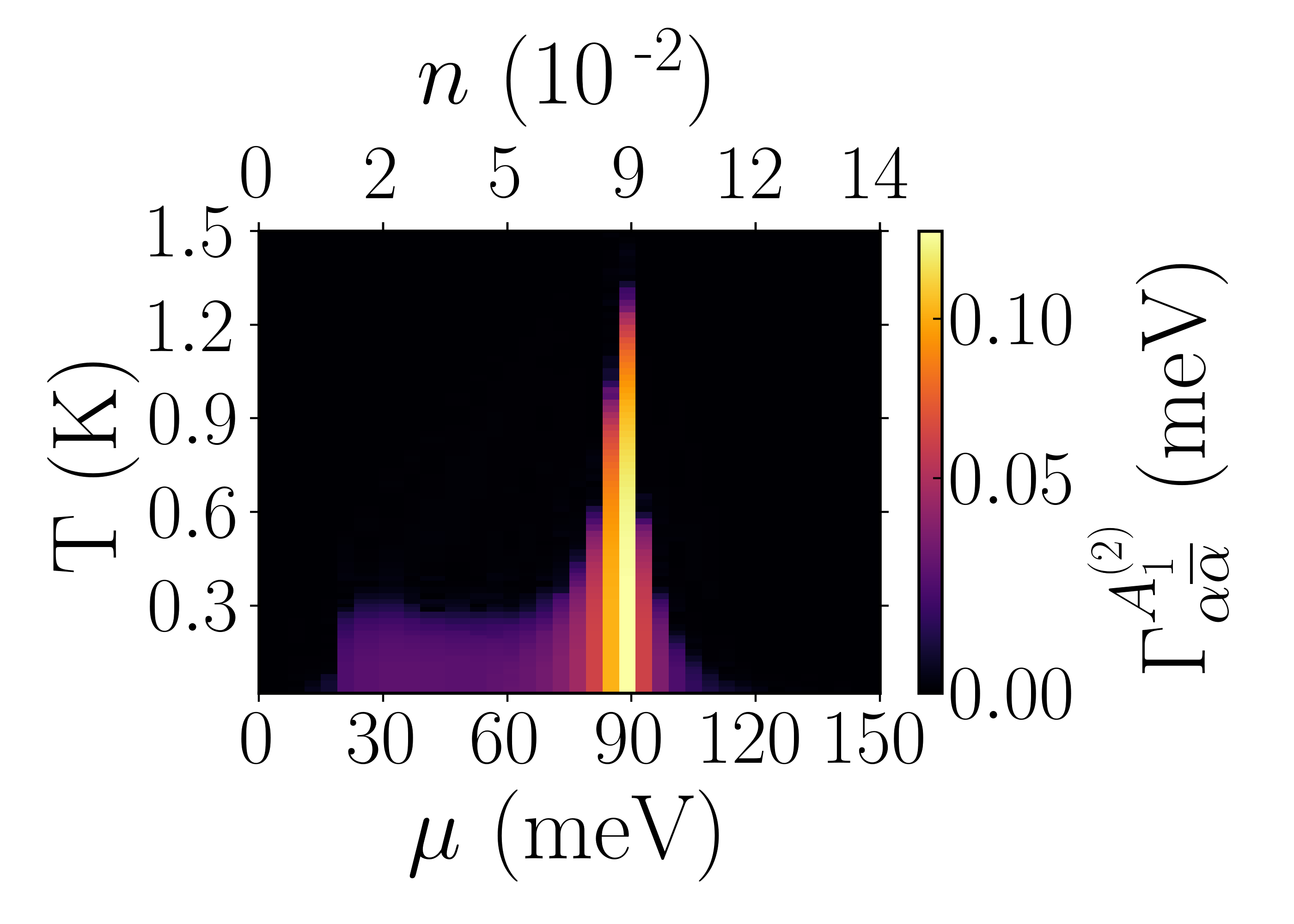}
    \put(-150,107){(c)}
    \caption{Magnitudes of symmetry-resolved order parameters in interlayer coupling channel ($J_{nnn} = J_{nnn}' = 0$) as a function of chemical potential $\mu$ (lower $x$ axis) and carrier concentration $n$ (upper $x$ axis) (a,b) for several values of the real-space pairing energy $J$, and (c) as a function of temperature for $J = 110$~meV. A clear peak in the pairing amplitudes (b) and critical temperature (c) corresponds to the van Hove singularity shown in Fig.~\ref{fig:dispersions}(d). In the plateau region ($\mu < 70$~meV), critical temperatures are in the range measured experimentally ($T_c \approx 0.3$~K). In panels (a) and (b), the red triangular points on the $\mu$ axis correspond to the values of the chemical potential chosen for further analysis.}
    \label{fig:NN}
\end{figure*} 
First, let us briefly analyze the electronic structure of 2DEG at the (111)-oriented LAO/STO interface. Fig.~\ref{fig:dispersions}(a,b) presents the dispersion relations in the normal state, determined along the $X$ and $Y$ directions of the Brillouin zone. As shown in panel (b), in the $Y$ direction, the lowest conduction band exhibits a large effective mass, with the maximum located around $85$~meV. This maximum gives rise to a van Hove singularity and the corresponding enhancement in the density of states (DOS) [Fig.~\ref{fig:dispersions}(d)], which may play a pivotal role in the superconducting properties of the system - according to BCS theory, the associated increase in the DOS enhances the stability of the superconducting state. A similar mechanism has recently been reported for the (001) LAO/STO interface in Ref.~\cite{Nakamura_001_2013}, where it manifests as an enhancement of the critical magnetic field. In contrast to the (001) direction, at the (111)-oriented LAO/STO interface, all three orbitals $d_{xy}$, $d_{yz}$, and $d_{xz}$ contribute equally to the Fermi surface, as shown in Fig.~\ref{fig:dispersions}(c). This indicates that the superconductivity in the (111) LAO/STO 2DEG is of multi-orbital character, and no significant difference between the orbital contributions to the superconducting state is expected.

In the following analysis, we vary the carrier density by tuning the chemical potential, which is defined as the energy measured from the bottom of the conduction band. To guarantee a single critical temperature for all orbitals, we set the Cooper pair-hopping energy at the level of $J' = 0.1J$ and $J_{nnn}' = 0.1J_{nnn}$.
We discuss the superconducting properties of the (111) LAO/STO interface under three different scenarios, in which the superconducting state is induced successively by: interlayer pairing between nearest neighbors (NN) (\ref{subsec:NN}), intralayer pairing between next-nearest neighbors (NNN) (\ref{subsec:NNN}), and a combination of both pairing types occurring simultaneously (NN+NNN) (\ref{subsec:Mix}). 

\subsection{NN pairing}
\label{subsec:NN}

We first focus on superconductivity mediated solely by the interlayer pairing, i.e. $J_{nnn} = J'_{nnn} = 0$. For clarity and to highlight the effects of the interlayer coupling, in the first approximation, we neglect the Hubbard interaction  by setting $U = V = 0$. A full discussion of the Hubbard term and its effect on the superconducting properties will be provided in Sec. \ref{subsec:HubbardInteraction}.

Figure~\ref{fig:NN} shows the magnitude of the symmetry-resolved superconducting order parameter as a function of chemical potential (carrier density) for several values of the pairing strength $J$. The electron density, marked on the upper axis, is defined as $n = \frac{1}{\sum_{l \alpha \sigma} 1} \sum_{l \alpha \sigma} n_{l \alpha \sigma}$, and determines the average number of electrons per site per spin orbital. Note that, in Fig.~\ref{fig:NN}, we present only the nonzero projections, which in the considered case correspond to the fully symmetric $A_1$ IR, typically associated with the extended $s$-wave pairing symmetry. The presence of two components within the $A_1$ IR results from the analysis of the superconducting gap in the full $\mathcal{T}$ module, which includes both spatial and orbital degrees of freedom, for which $N_{A_1} = 2$. In this case, the superconducting order parameter comprises two linearly independent components, distinguishing whether the direction of pairing in real space — expressed as the anisotropy of the pairing amplitude to neighboring sites — aligns with the orientation of a given orbital. 
Projection onto $A_{1}^{(1)}$ quantifies the superconducting coupling when the direction of real-space pairing aligns with that of a specific orbital, whereas $A_{1}^{(2)}$ corresponds to the case when they are not aligned. It is important to emphasize that this specific relation between the pairing amplitudes and the orbital orientations in real space, within the $A_{1}^{(2)}$ projection, leads to a maximization of pairing along the orbital direction in wave vector space - see further discussion below. 
To better illustrate this dependence, consider the pairing amplitude $\Gamma_{d_{xy}} ( \vec{\delta}_{nn}^{(1)})$, which is aligned with the orientation of the $d_{xy}$ orbital oriented along $\vec{\delta}_{nn}^{(1)}$, and thus contributes primarily to the $A_{1}^{(1)}$ representation. In contrast, the amplitude $\Gamma_{d_{xy}} ( \vec{\delta}_{nn}^{(2)} )$, whose direction is determined by $\vec{\delta}_{nn}^{(2)}$, is not aligned with the $d_{xy}$ orbital orientation and contributes primarily to $A_{1}^{(2)}$ projection.
\begin{figure}[!t]
    \centering
    \includegraphics[width=1\linewidth]{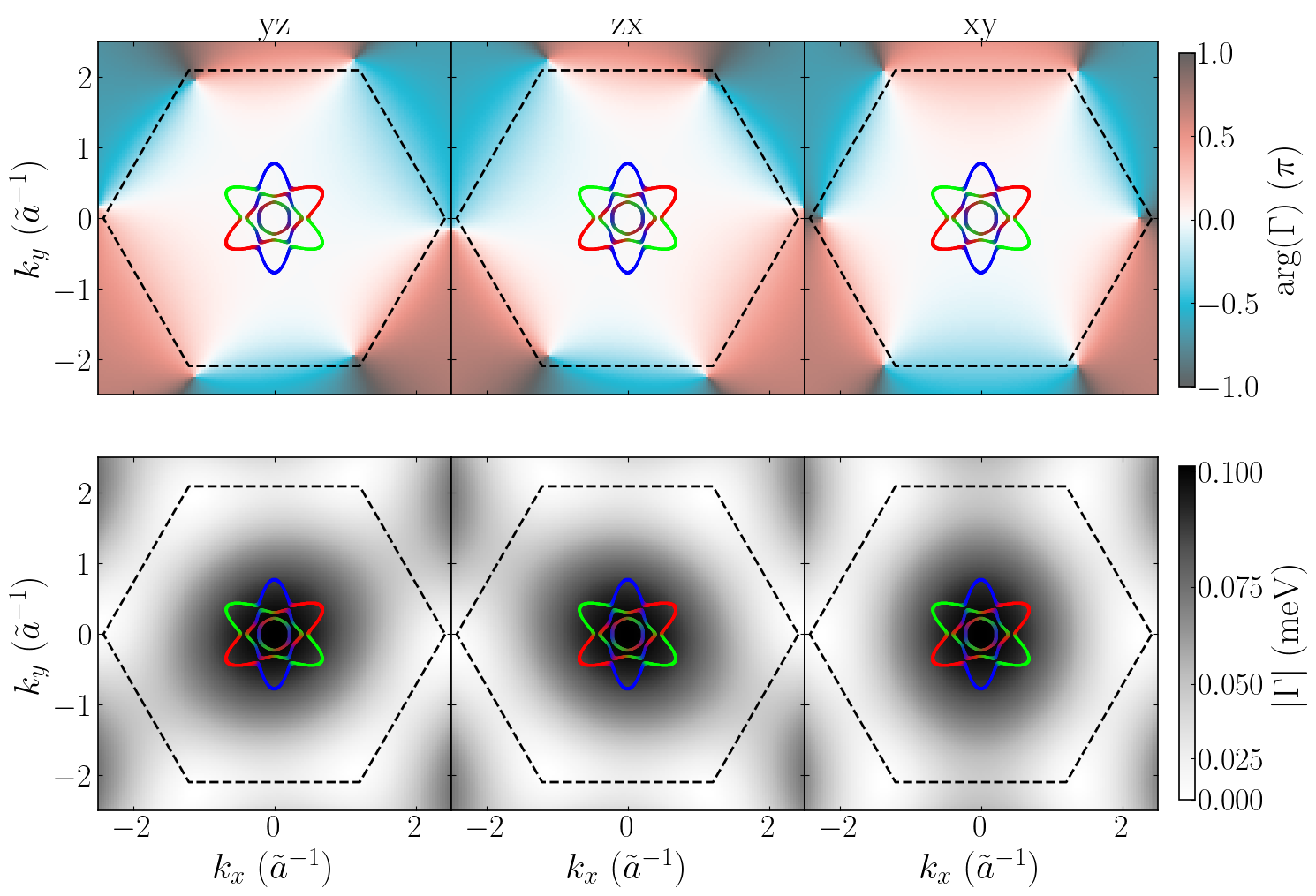}
     \put(-250,155){(a)}
     \vspace{0.5cm}\\
    \includegraphics[width=1\linewidth]{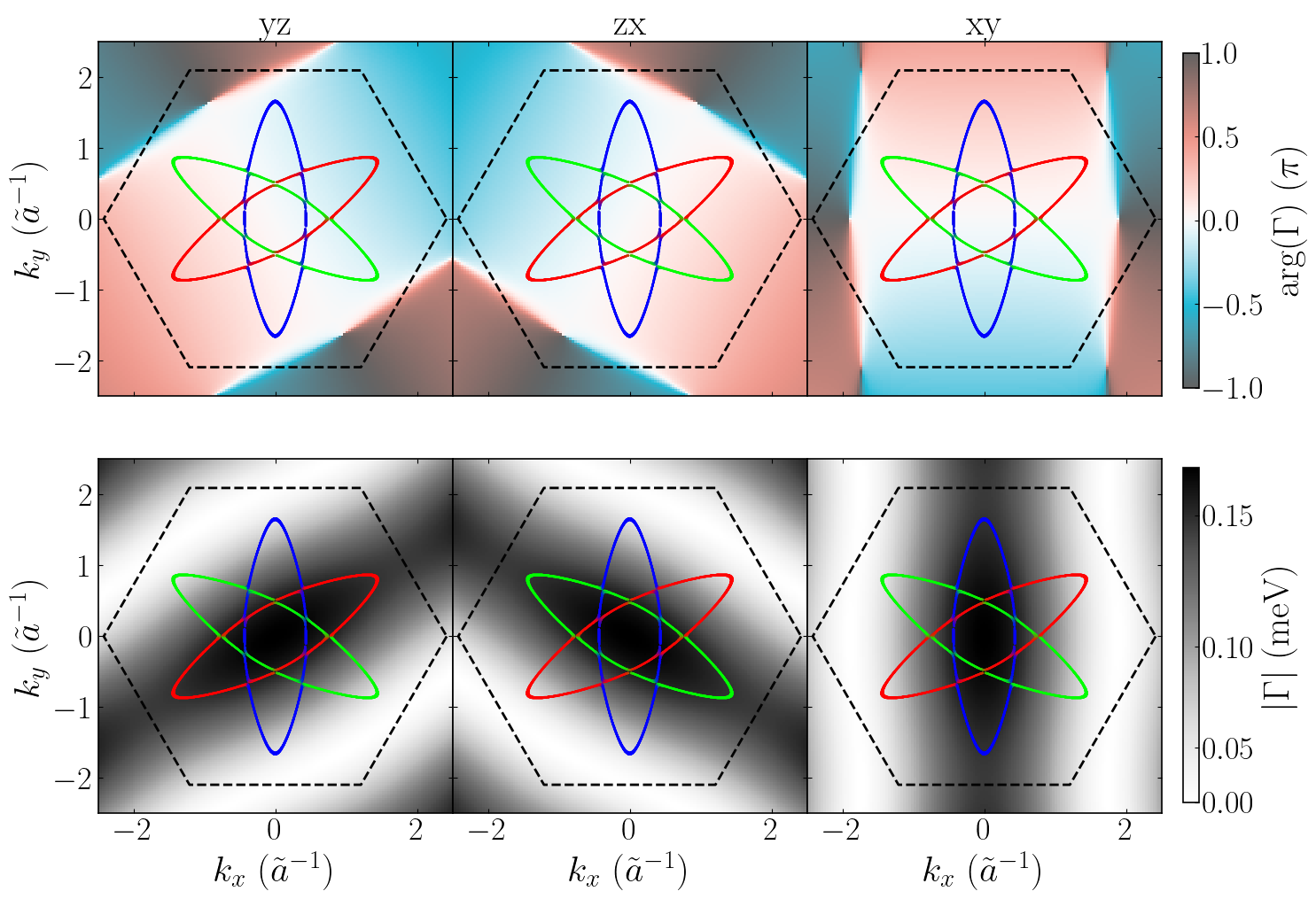}
     \put(-250,155){(b)}
     \vspace{0.5cm}\\
    \includegraphics[width=1\linewidth]{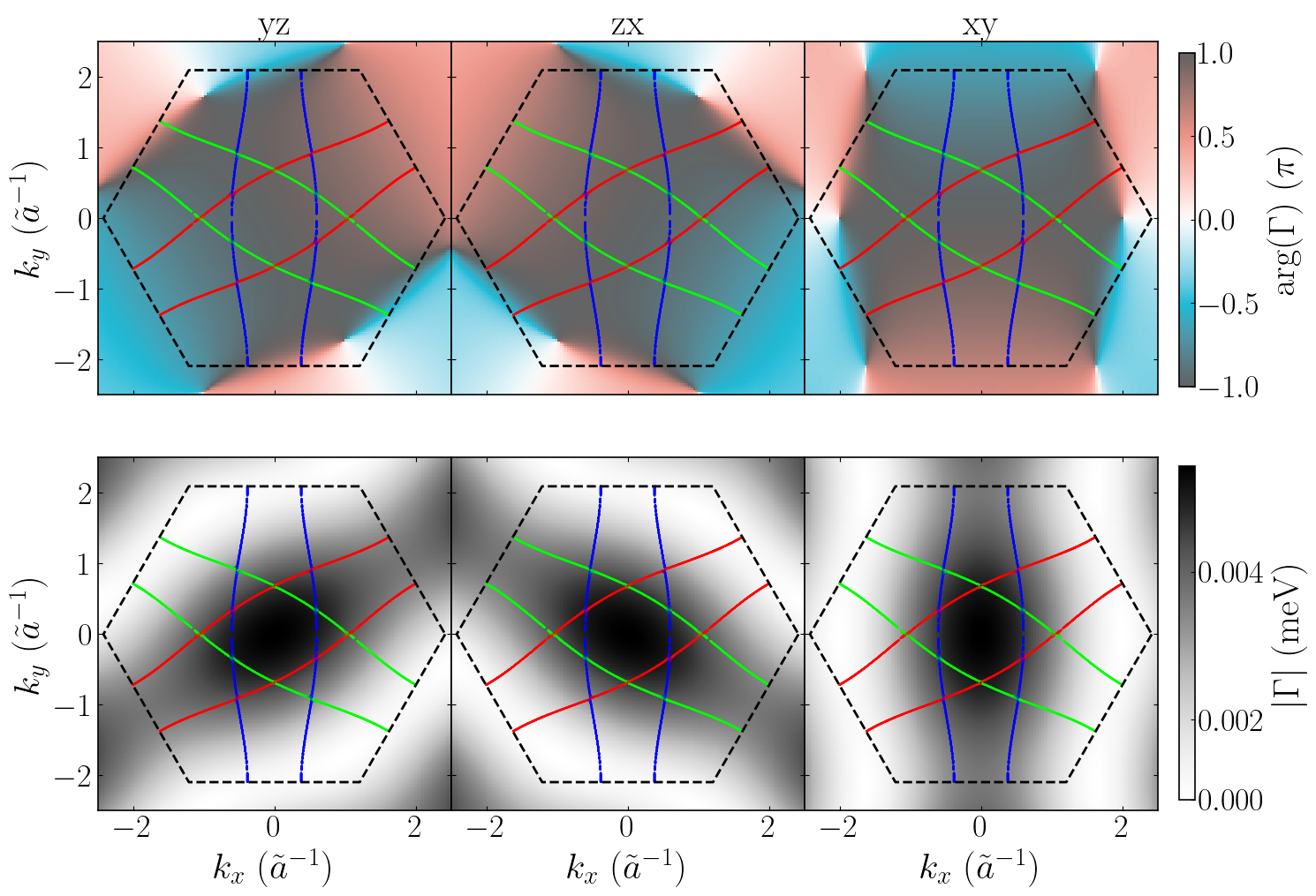}
     \put(-250,155){(c)}
    \caption{Maps of $\Gamma_{l\alpha \overline{\alpha}} \left ( \kvec \right )$ [Eq. \eqref{eq:gamma_nn}] presented in the form of its module (lower row) and the argument (upper row),
    for $t_{2g} = \{d_{yz}, d_{zx}, d_{xy} \}$ orbitals presented in consecutive columns. Results for (a) $\mu= 31$~meV, (b) $\mu= 79$~meV and (c) $\mu= 141$~meV - see Fig.~\ref{fig:NN}. Each panel displays the first Brillouin zone, indicated by gray dotted lines, along with the corresponding Fermi surface. In the latter, orbital-resolved contributions from the $d$-orbitals are represented by color: red for $d_{yz}$, green for $d_{zx}$, and blue for $d_{xy}$.}
    \label{fig:NN_GammaK}
\end{figure}

\begin{figure}[!t]
    \includegraphics[width=0.5\linewidth]{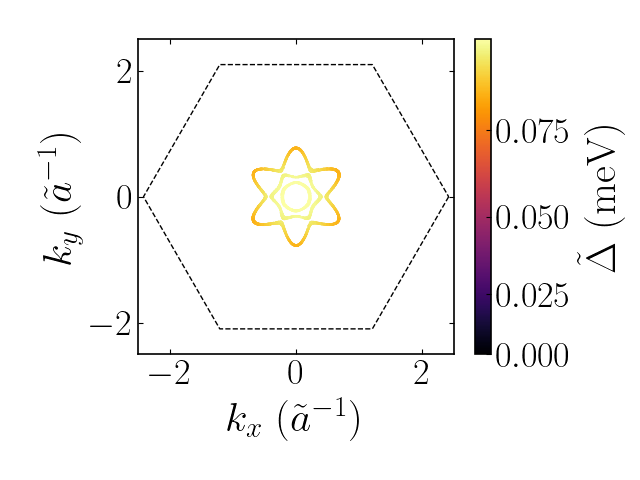}
    \put(-120,75){(a)}
    \includegraphics[width=0.5\linewidth]{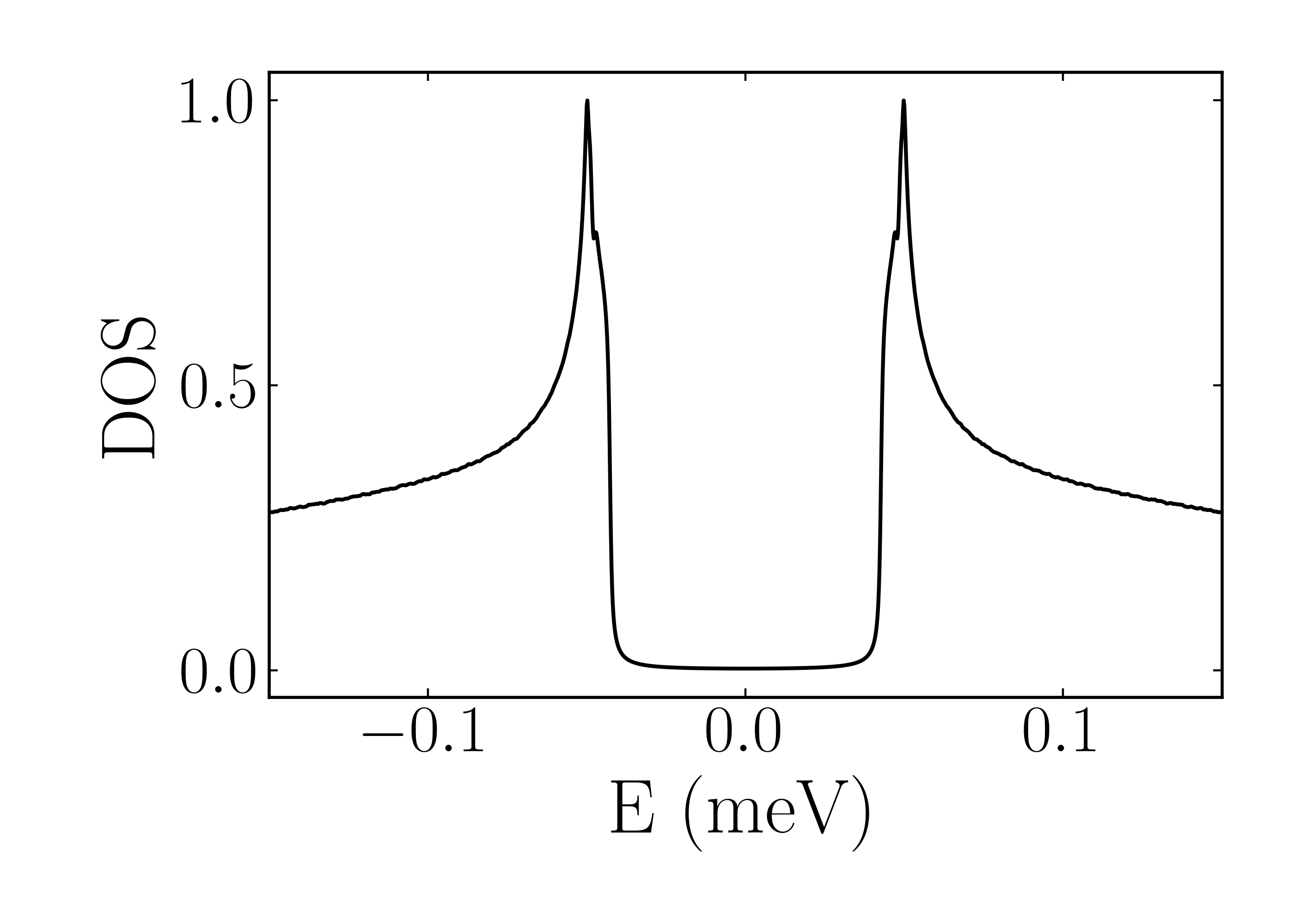}
    \put(-120,75){(b)}
    \caption{(a) Effective superconducting gap at the Fermi surface and (b) the resulting density of states (DOS). Results for $\mu = 31$~meV and the interlayer pairing energy $J = 110$~meV.}
    \label{fig:NN_Dos_Gap}
\end{figure}

The foregoing analysis provides strong evidence for an intrinsic \textit{spatial} (in the $\mathcal{R}$ module) dependence of the real-space pairing amplitude associated with a given orbital.  This can be understood by simply examining the Fermi surface shown in Fig.~\ref{fig:dispersions}(c). Since our model includes only intraorbital pairing, we can consider a Cooper pair formed in the outermost band along the $k_y$ direction, which requires both electrons of the pair to originate from the $d_{xy}$ orbital. In the framework of the solely spatial symmetry ($\mathcal{R}$), we can study how the pairing between electrons changes under rotation of momentum, by the angle defined by the crystalline symmetry, assuming that both electrons remain in the same orbital. However, if we rotate the momenta clockwise by $\pi /3$, the pairing in the $d_{xy}$ orbital becomes irrelevant, as the Fermi surface in this direction is primarily determined by the $d_{yz}$ orbital. This implies that the spatial component of the pairing amplitude within a given orbital exhibits a pronounced directional (anisotropic) character. Note, however, that it does not necessarily give rise to the exotic symmetry of the total superconducting gap, as the full symmetry must be evaluated by accounting for the multiband nature of the system with defined symmetries of the orbitals.

\begin{figure*}[!t]
\includegraphics[width=0.33\linewidth]{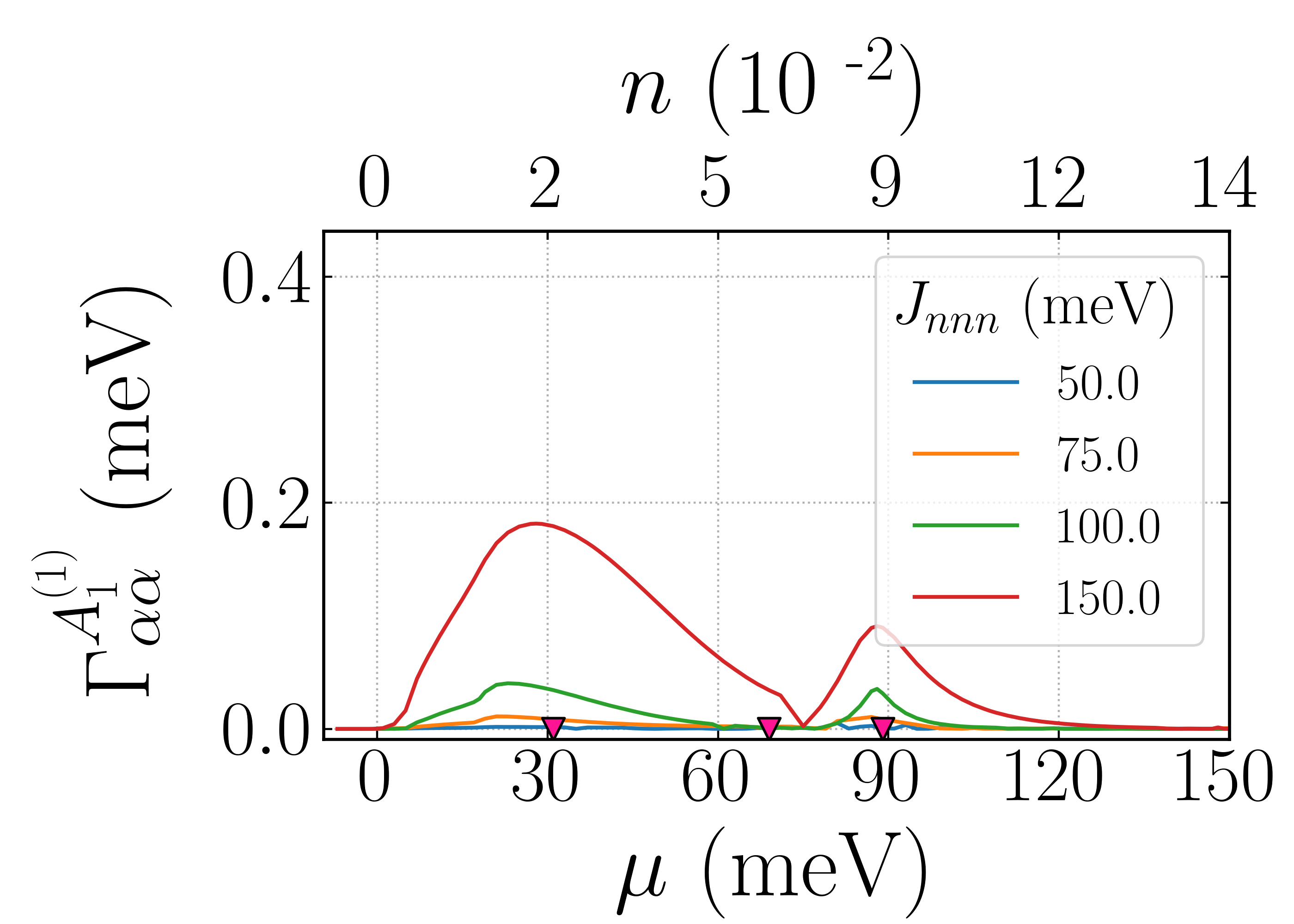}
\put(-150,110){(a)}
\includegraphics[width=0.33\linewidth]{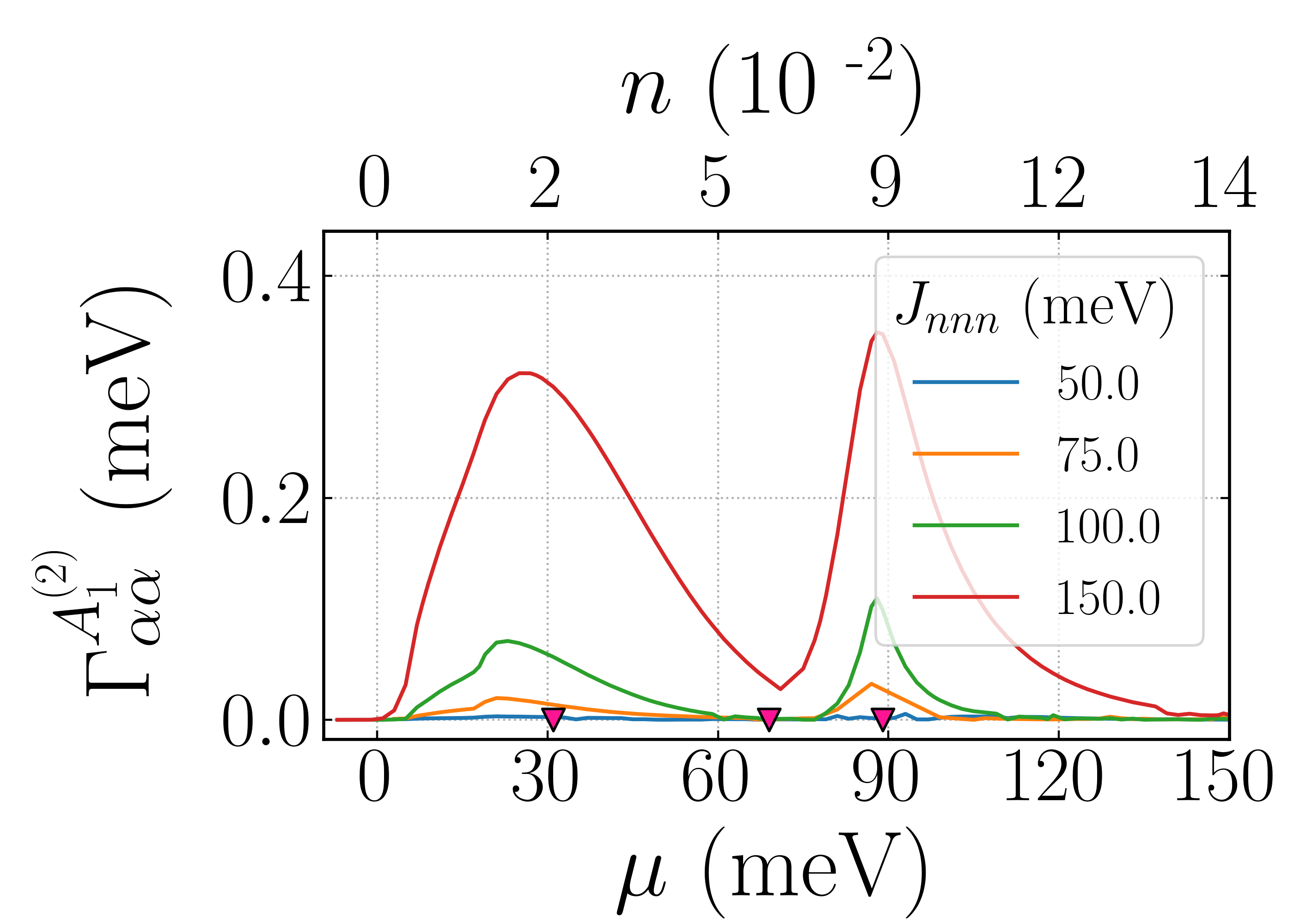}
\put(-155,110){(b)}
\includegraphics[width=0.33\linewidth]{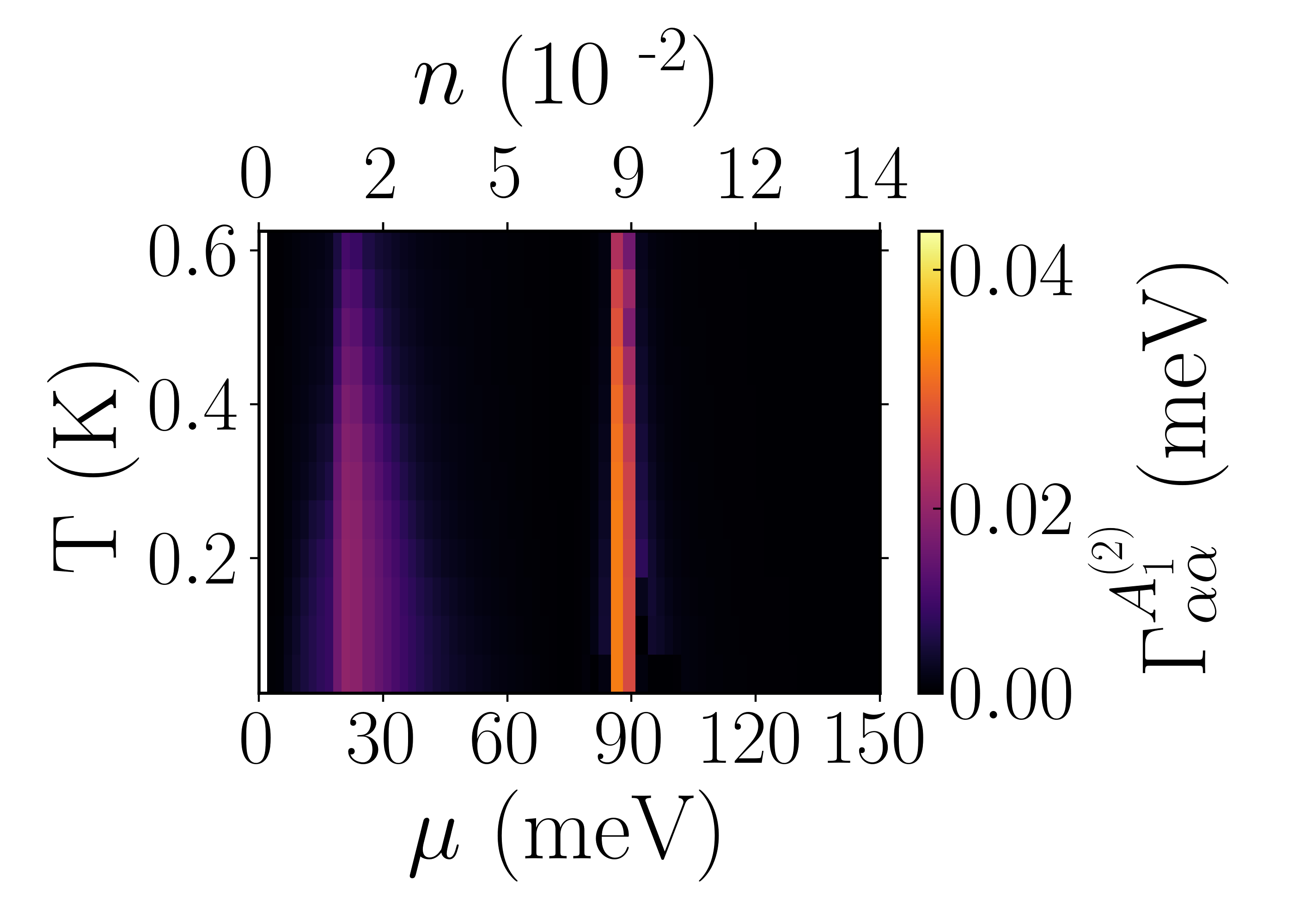}
\put(-150,110){(c)}
\caption{
Magnitudes of symmetry-resolved order parameters in intralayer coupling channel ($J = J' = 0$) as a function of chemical potential $\mu$ (lower $x$ axis) and carrier concentration $n$ (upper $x$ axis) (a,b) for several values of real-space pairing energy $J_{nnn}$, and (c) as a function of temperature for $J_{nnn} = 75$~meV. In panels (a) and (b), the red triangular points on the $\mu$ axis correspond to the values of the chemical potential chosen for further analysis.}
\label{fig:NNN}
\end{figure*}

Figure~\ref{fig:NN} reveals that superconductivity emerges at low chemical potentials, specifically around $\mu \approx 5$~meV, and is characterized by the activation of both $A_1$ IR components. The pairing channel corresponding to the $A_1^{(1)}$ projection—associated with real-space coupling along directions aligned with the orbital lobes—is significantly suppressed and vanishes in the vicinity of the van Hove singularity. In contrast, the $A_1^{(2)}$ component, which corresponds to pairing in directions misaligned with the orbital axes, exhibits a marked enhancement near the singularity, attaining a global maximum before diminishing at higher chemical potentials.

The evolution of the superconducting order parameter within both $A_1^{(1)}$ and $A_1^{(2)}$ channels can be tracked in Fig.~\ref{fig:NN_GammaK}, where we present momentum-space maps of the pairing amplitude $\Gamma_{l \alpha \overline{\alpha}}(\kvec)$ [as defined in Eq.\eqref{eq:gamma_nn}] for each $d$ orbital and three selected values of the chemical potential, indicated in Fig.~\ref{fig:NN}. At low chemical potential [Fig.~\ref{fig:NN_GammaK}(a)], the pairing amplitude is predominantly real and nearly isotropic for all orbitals. Since the Fermi surface is centered near the $\Gamma$ point of the Brillouin zone, the pairing in each orbital lacks any pronounced directional dependence associated with the orbital geometry. This isotropy results in comparable contributions from both $A_1^{(1)}$ and $A_1^{(2)}$ components at the onset of superconductivity, as shown in Fig.~\ref{fig:NN}.
As the chemical potential increases [Fig.~\ref{fig:NN_GammaK}(b)], the spatial asymmetry of each orbital’s pairing becomes more pronounced, exhibiting a larger maximum along a particular direction. For each orbital, $ | \Gamma_{l \alpha \overline{\alpha}} ( \kvec ) |$ has a clear maximum along the orientation of the orbital. This is in line with the vanishing of the $A_1^{(1)}$ component and the enhancement in $A_1^{(2)}$ presented in Fig.~\ref{fig:NN}. At the van Hove singularity, the anisotropy of the orbital-resolved pairing amplitudes becomes fully aligned with the highly directional character of the Fermi surface.
This results in a suppression of the $A_1^{(1)}$ projection and a concurrent enhancement of the $A_1^{(2)}$ component.  Finally, in Fig.~\ref{fig:NN_GammaK}(c) we can see that for high chemical potentials, well above the van Hove singularity, the magnitude of $ | \Gamma_{l \alpha \overline{\alpha}} ( \kvec ) |$  is strongly suppressed (refer to the scale on the right). Simultaneously, the Fermi surface spreads away from the region of maximal pairing strength in reciprocal space, resulting in vanishing of superconducting properties, as shown in Fig.~\ref{fig:NN}. It is also important to note that regardless of the chemical potential, the modulus of the pairing amplitude is strictly periodic within the Brillouin zone. This periodicity ensures that the superconducting quantities preserve the full $C_{6v}$ symmetry, as expected from the underlying lattice structure.
Here, we should point out that Fig.~\ref{fig:NN_GammaK} presents results for an arbitrary layer, while for classification in the $C_{6v}$ point group, both layers are taken into account, by simply adding their contributions. The latter approach leads to strictly real components, as in the spin-singlet pairing scenario the inversion of lattice index $\alpha \overline{\alpha} \rightarrow \overline{\alpha} \alpha$ leads to the conjugation of order parameter $\Gamma_{l \alpha \overline{\alpha}} (\kvec ) = ( \Gamma_{l \overline{\alpha} \alpha} (\kvec ))^*$ - see Eq.~(\ref{eq:gamma_nn}).

In Fig.~\ref{fig:NN}, we demonstrate that for a considered range of real-space pairing strengths $J$, the qualitative behavior of the superconducting order parameter remains unchanged. Note that $J$ in our model can be used as a tuning parameter to adjust observable physical quantities—such as the critical temperature $T_c$ or the superconducting gap at the Fermi surface — to match experimentally observed values. As shown in Fig.~\ref{fig:NN}(c), we successfully reproduce $T_c \approx 300$~mK, in agreement with experimental measurements, at low carrier concentrations for a pairing strength of $J=110$~meV. Achieving lower critical temperatures necessitates a finer integration grid, which substantially increases the computational cost. Therefore, in the following analysis, we focus on the general features of the superconducting phase diagram, arguing that these can be scaled by tuning the pairing strength $J$ to reproduce experimental observations.

Upon obtaining self-consistent pairing amplitudes, we diagonalize the Hamiltonian \eqref{eq:H_Nambu} to investigate the effective energy gap at the Fermi surface, which we denote as $\tilde{\Delta} ( \kvec )$. The results for $\mu = 31$~meV are presented in Fig.~\ref{fig:NN_Dos_Gap}(a), in which we can observe that the effective superconducting gap is nodeless and has a perfect $C_{6v}$ symmetry, as expected from a fully symmetric $A_1$ IR corresponding to the extended $s$-wave symmetry. Despite the extended $s$-wave symmetry is nearly isotropic in the vicinity of the $\Gamma$ point, the specific Fermi surface topology imposed by the hexagonal symmetry results in a clear angular dependence of the superconducting gap across each band. It should be noted that such behavior shall, in principle, be observable in tunneling conductance measurements, as it changes the density of states (DOS) around the gap. As shown in Fig.~\ref{fig:NN_Dos_Gap}(b), the angular dependence leads to a slight smoothing of the coherence peaks, deviating from the sharp $U$-shaped profile typically associated with conventional $s$-wave symmetry and isotropic Fermi surface. Here, we emphasize that this behavior is an inherent feature of the underlying $C_{6v}$ symmetry of the Fermi surface.

\begin{figure}[!htp]
    \centering
    \includegraphics[width=0.5\linewidth]{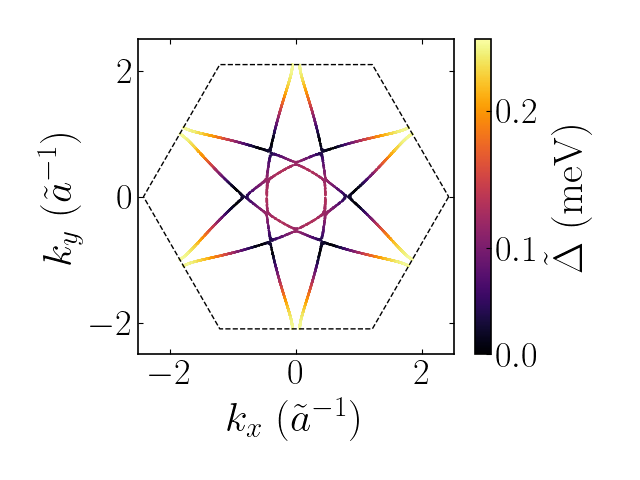}
    \put(-120,75){(a)}
    \includegraphics[width=0.5\linewidth]{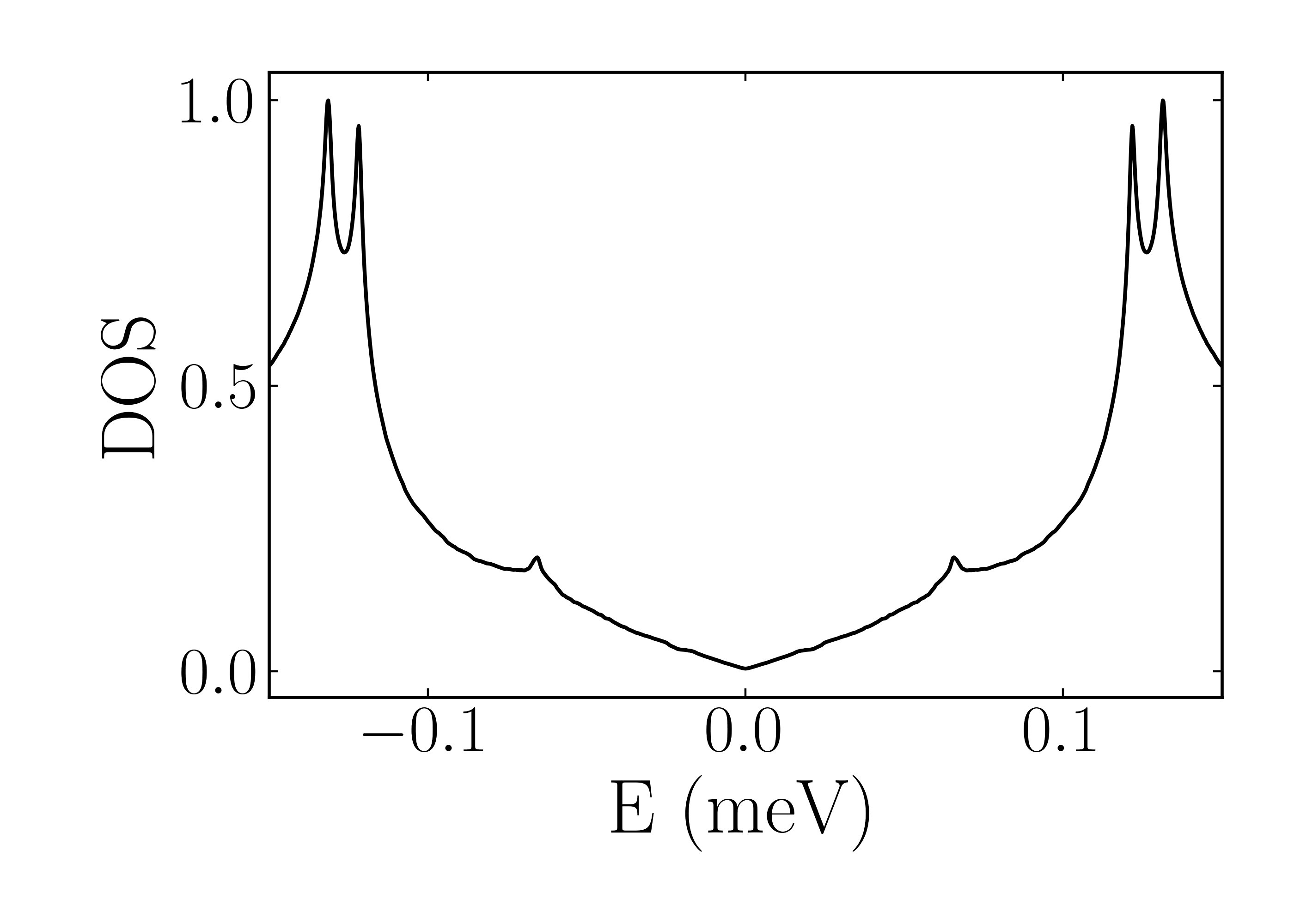}
    \put(-120,75){(b)}
    \caption{(a) Effective superconducting gap at the Fermi surface and (b) the resulting density of states (DOS). Results for $\mu = 89$~meV corresponding to the second dome in Fig.~\ref{fig:NNN} and the interlayer pairing energy $J_{nnn} = 75$~meV. 
    }
    \label{fig:NNN_Dos_Gap}
\end{figure}

\subsection{NNN pairing}
\label{subsec:NNN}

Let us proceed to perform a similar analysis for the intralayer coupling scenario. Now, we set the nearest-neighbor pairing $J = J' = 0$ and interorbital coupling $J_{nnn}\ne 0$ while  $J'_{nnn} = 0.1J_{nnn}$. Again, to underline superconducting properties related to that specific scenario, we set the Hubbard interaction energies $U=V=0$.

\begin{figure}[!htp]
    \centering
    \includegraphics[width=1\linewidth]{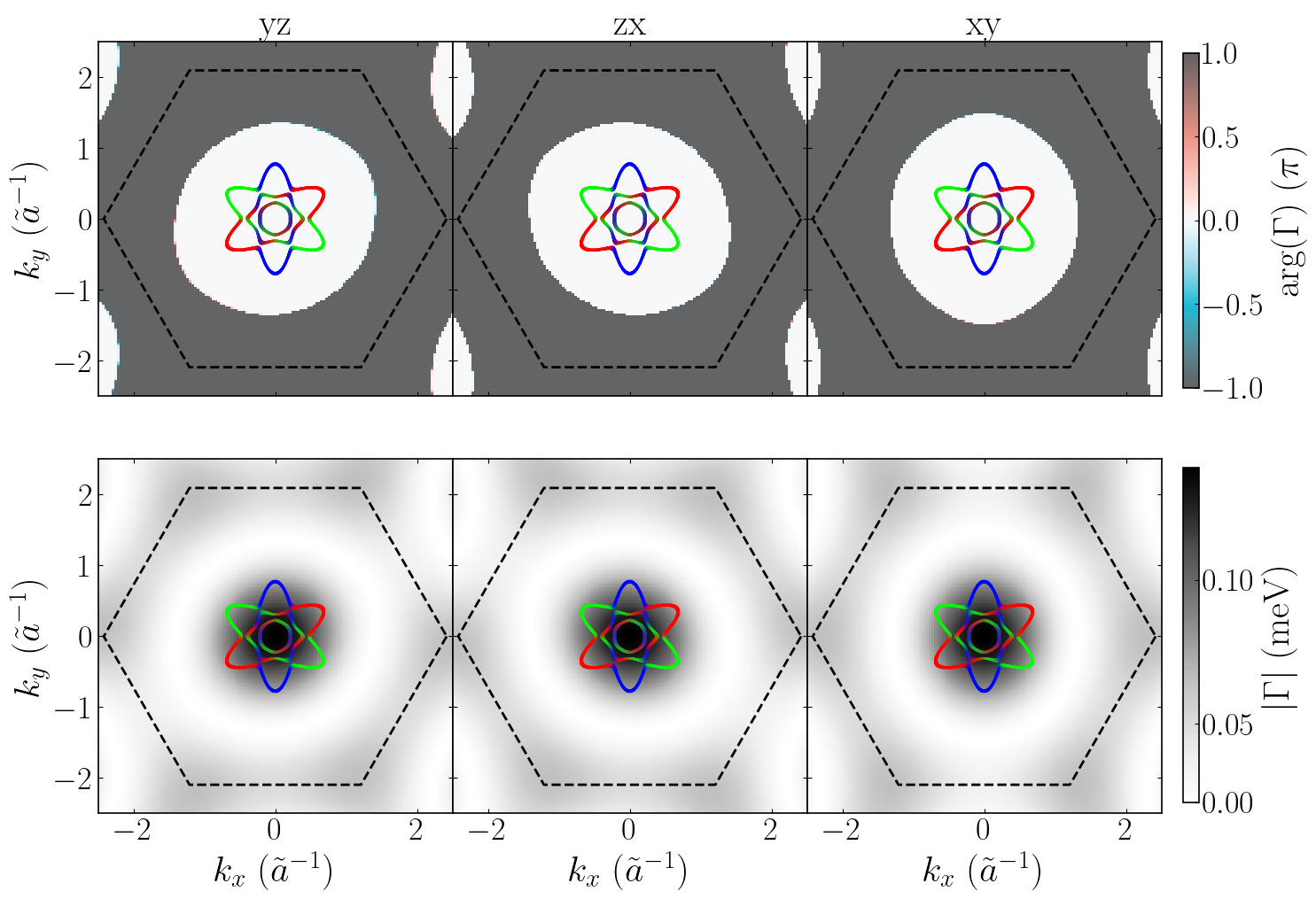}
    \put(-250,155){(a)}
    \vspace{0.5cm}\\
    \includegraphics[width=1\linewidth]{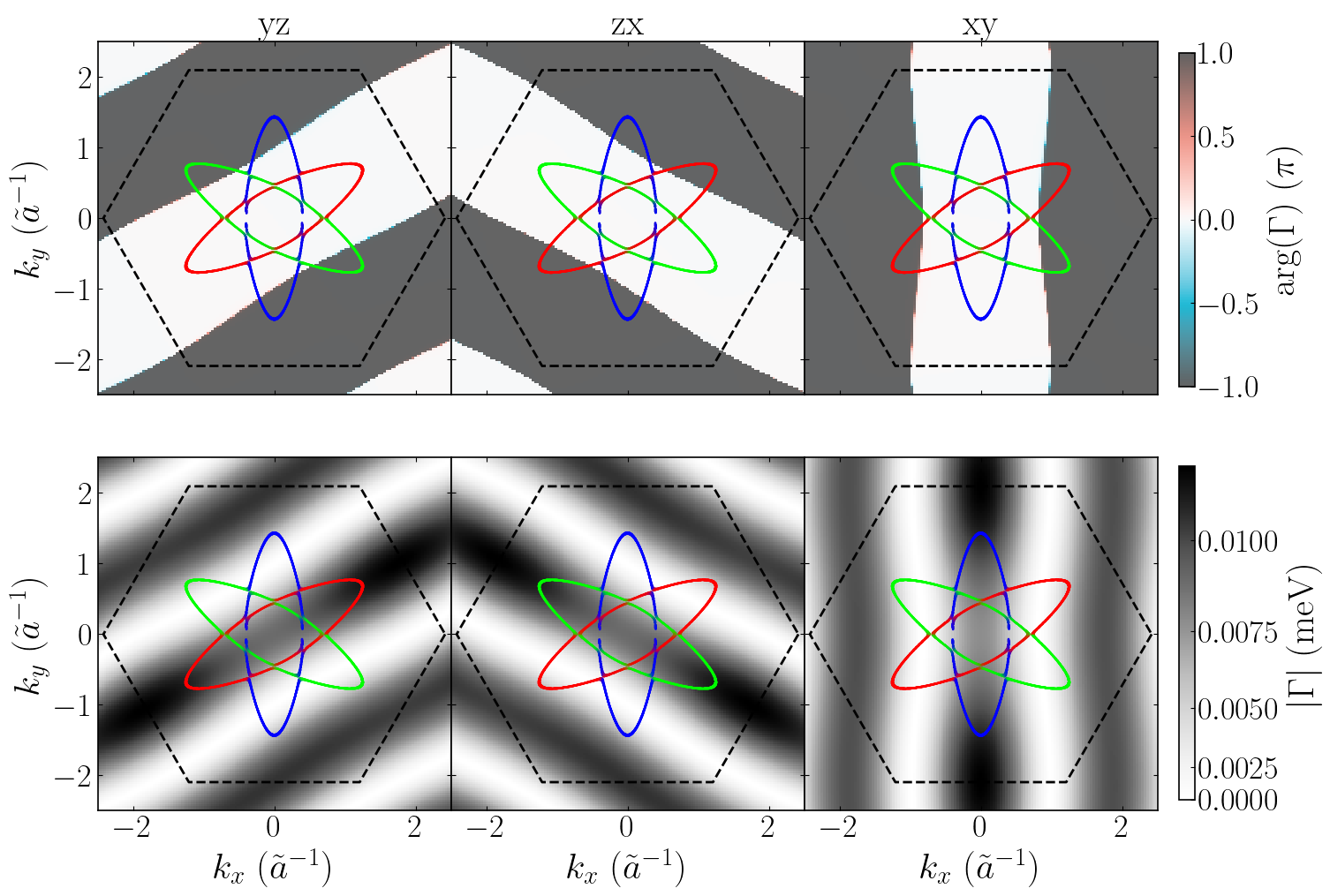}
    \put(-250,155){(b)}
    \vspace{0.5cm}\\
    \includegraphics[width=1\linewidth]{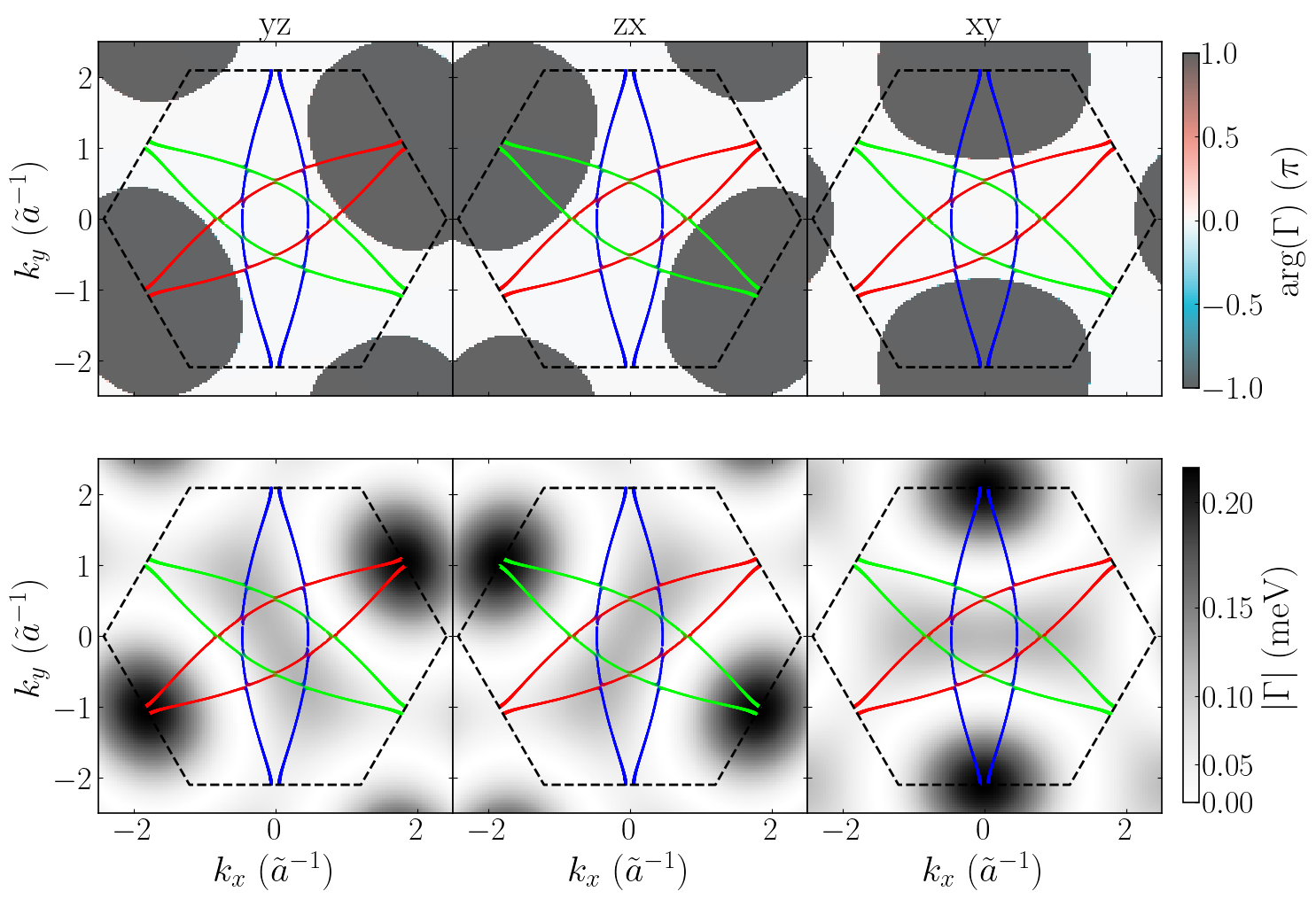}
    \put(-250,155){(c)}
    \vspace{0.5cm}\\
    \caption{Maps of $\Gamma_{l\alpha \alpha} \left ( \kvec \right )$ [Eq. \eqref{eq:gamma_nnn}] presented in the form of its module (lower row) and the argument (upper row),
    for $t_{2g} = \{d_{yz}, d_{zx}, d_{xy} \}$ orbitals presented in consecutive columns. Results for (a) $\mu= 31$~meV, (b) $\mu= 69$~meV and (c) $\mu= 89$~meV - see Fig.~\ref{fig:NNN}. Each panel displays the first Brillouin zone, indicated by gray dotted lines, along with the corresponding Fermi surface. In the latter, orbital-resolved contributions from the $d$-orbitals are represented by color: red for $d_{yz}$, green for $d_{zx}$, and blue for $d_{xy}$.}
    \label{fig:NNN_GammaK}
\end{figure}

Fig.~\ref{fig:NNN} displays the symmetry-resolved superconducting order parameter as a function of the chemical potential (carrier density) for different $J_{nnn}$. In this case, pairing in the two Ti layers is considered separately — see Eq.~(\ref{eq:gamma_nnn}) — where, in each layer, a single node is coupled to the six neighboring sites - red and blue points in Fig.~\ref{fig:hexagonal_111}.
These layers are not equivalent due to the breaking of inversion symmetry near the interface and the corresponding perpendicular electric field [Eq.~(\ref{eq:H_externalElectric})], which results in the potential $v$ that is opposite in the two layers. Consequently, one of the layers becomes energetically preferred, which leads to a greater contribution from that layer to the overall carrier density and DOS. For clarity, we present results only for the layer that lies higher in energy, as it generates a smaller magnitude of the order parameter. Simultaneously, we have verified that the qualitative behavior of the superconducting order parameter in both layers is the same, differing only in magnitude.

\begin{figure*}[!htp]
\centering
\includegraphics[width=0.25\linewidth]{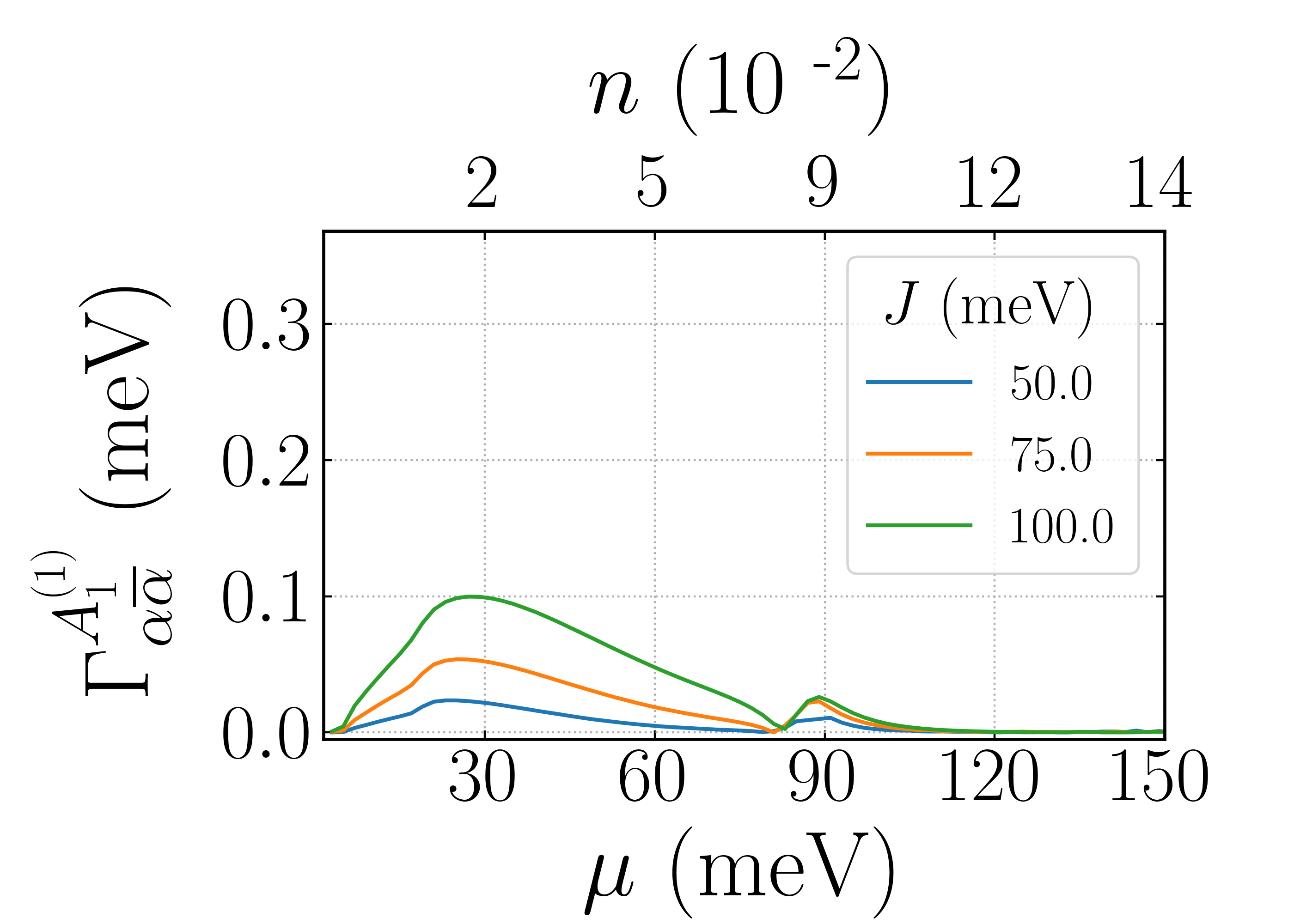}
\put(-110,80){(a)}
\includegraphics[width=0.25\linewidth]{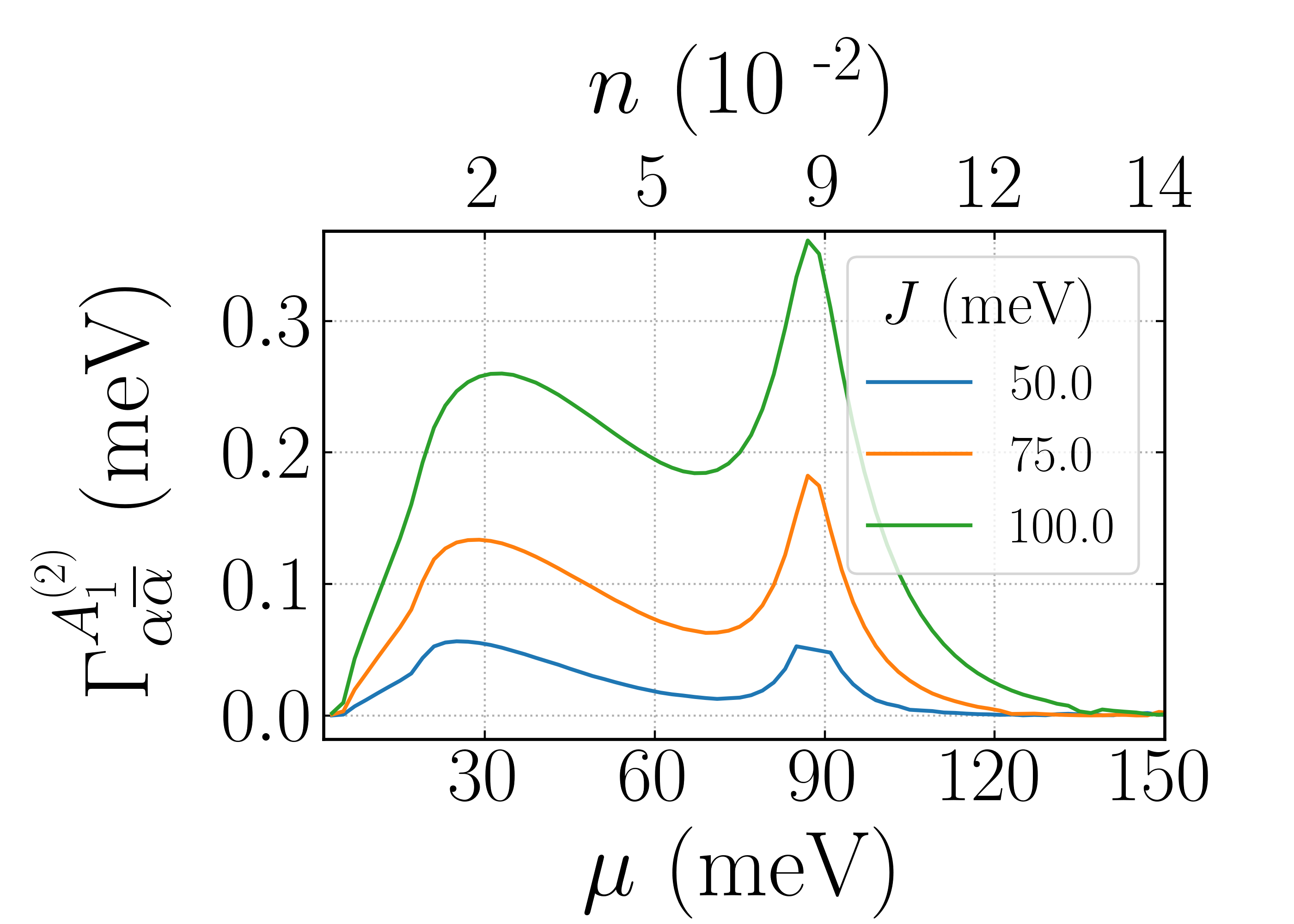}  
\put(-110,80){(b)}
\includegraphics[width=0.25\linewidth]{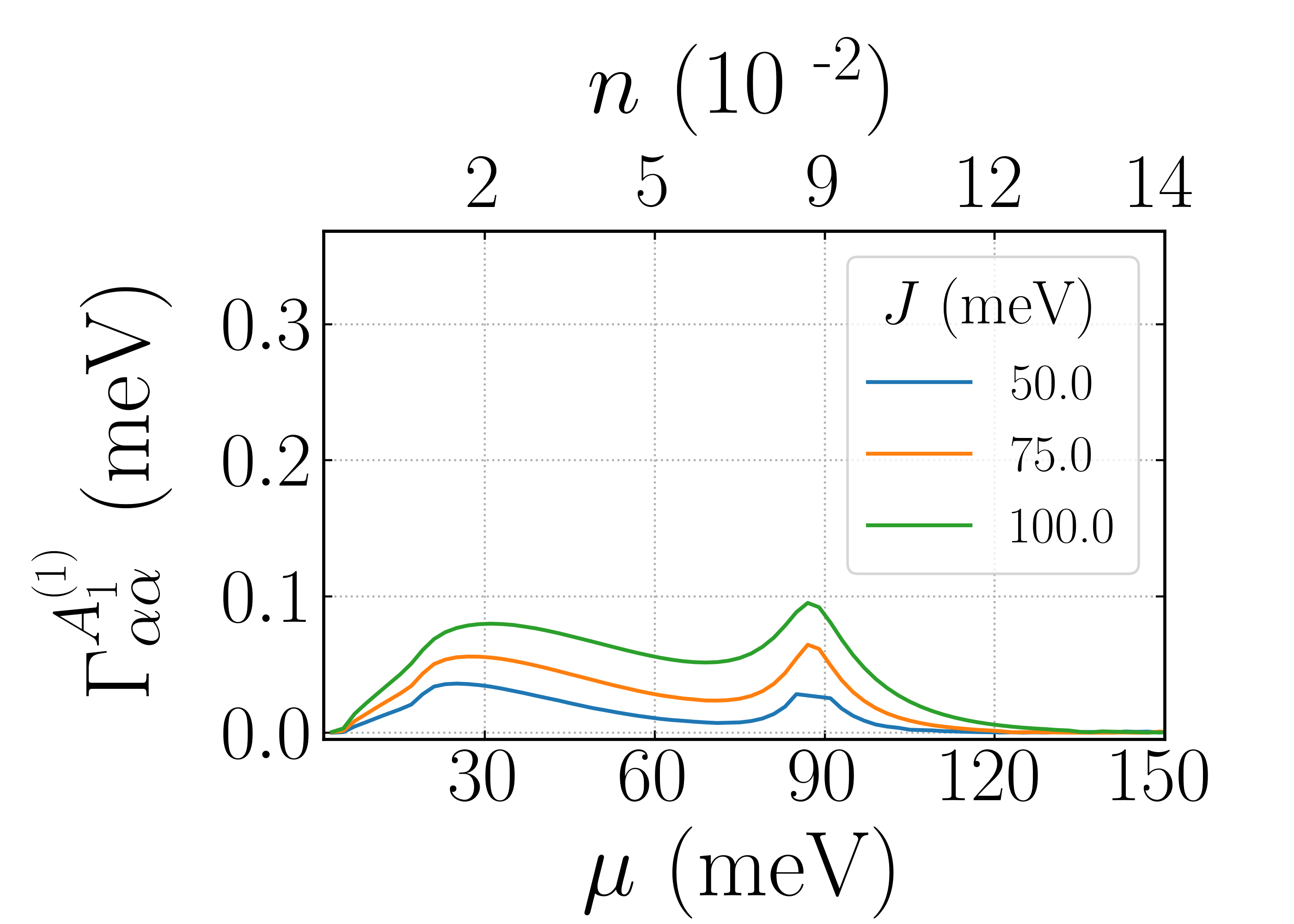}
\put(-110,80){(c)}
\includegraphics[width=0.25\linewidth]{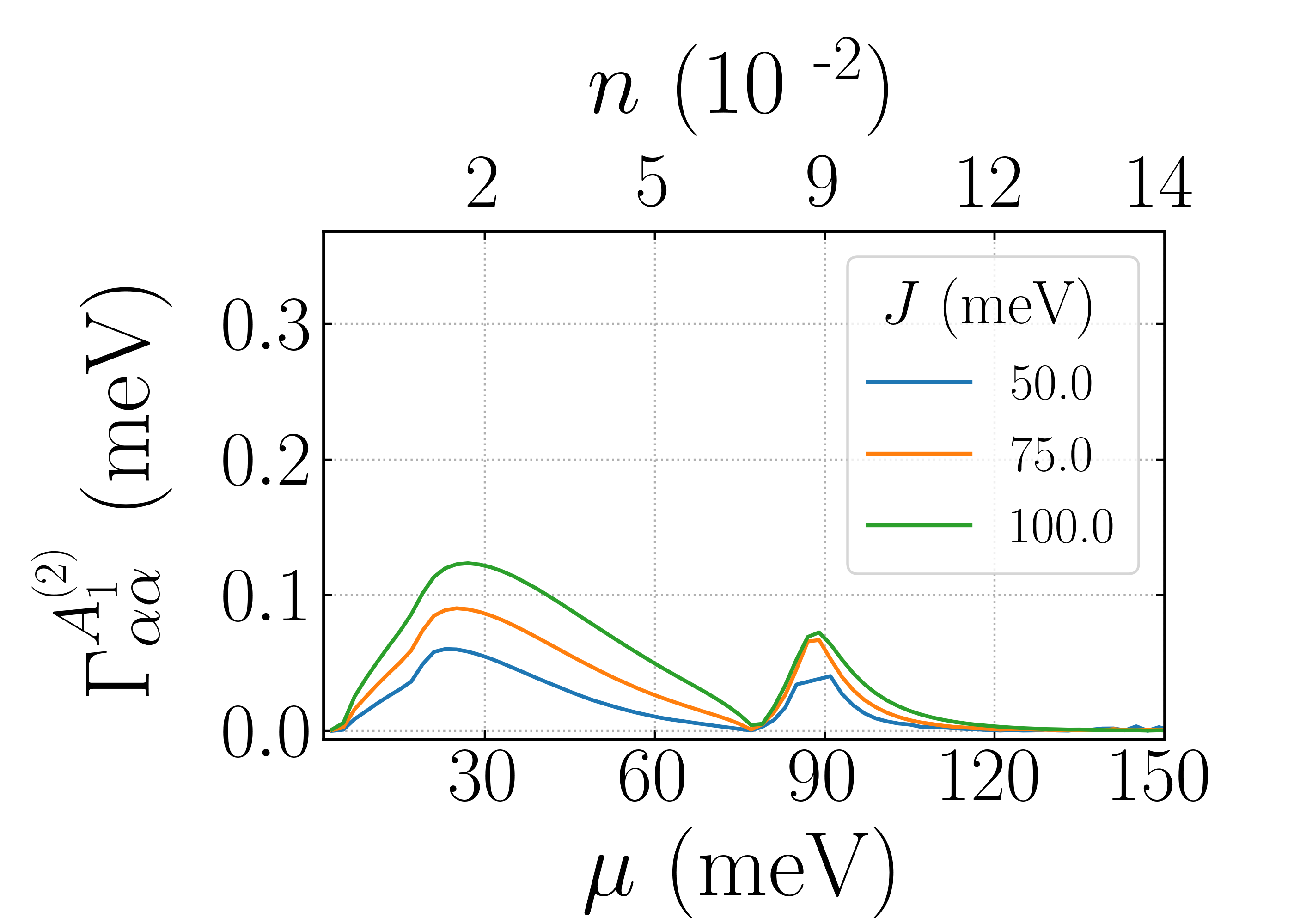}  
\put(-110,80){(d)}

\caption{Magnitudes of symmetry-resolved order parameters in inter- (a,b) and intralayer (c,d) coupling channel as a function of chemical potential $\mu$ (lower $x$ axis) and carrier concentration $n$ (upper $x$ axis) for several values of real-space pairing energy $J$ and fixed $J_{nnn} = 75$~meV.}
\label{fig:mix}
\end{figure*}

Figure~\ref{fig:NNN} clearly demonstrates that the superconducting properties in the intralayer pairing scenario differ significantly from those in the interlayer pairing case — compare with Fig.~\ref{fig:NN}. Superconductivity emerges at even lower carrier concentrations and increases rapidly, reaching a global maximum for both $A_1$ projections at $\mu \approx 30$~meV. This value of chemical potential corresponds to the carrier density $n \approx 8 \times 10^{13}$~cm\textsuperscript{-2}, which is comparable to the range reported in experiments \cite{Monteiro2019}. In this range, the estimated critical temperature $T_c \approx 600$~mK for $J=75$~meV - see Fig.~\ref{fig:NNN}(c).  As the chemical potential increases, superconductivity nearly disappears at intermediate carrier concentrations and then reemerges near the van Hove singularity, resulting in the characteristic double-dome structure. Notably, in the considered pairing scenario, even lower $J_{nnn}$ values generate a larger order parameter in the low electron density regime compared to the interlayer pairing. This, together with the appearance of the superconducting dome commonly observed in experiments \cite{Rout2017}, suggests the significance of NNN intralayer pairing in the superconducting phase diagram, potentially making it the dominant pairing mechanism in the (111) LAO/STO interface. 

Differences between the interlayer and intralayer pairing scenarios are also evident in the effective energy gap at the Fermi surface and the corresponding DOS. Although the shape of DOS for the low-energy dome is qualitatively similar to that found in the NN pairing case [Fig.~\ref{fig:NN_Dos_Gap}], we observe a strikingly different behavior in the range of the electron concentration (chemical potential) related to the second dome. In Fig.~\ref{fig:NNN_Dos_Gap}(a) we present the effective superconducting gap $\tilde{\Delta} ( \kvec )$ at the Fermi surface, along with the corresponding DOS in Fig.~\ref{fig:NNN_Dos_Gap}(b), both calculated for $\mu = 89$~meV. We observe that the superconducting gap vanishes at certain points on the Fermi surface, repeated according to the $C_{6v}$ symmetry. As a consequence, the DOS exhibits the characteristic $V$-shaped behavior, typical for order parameters exhibiting nodal lines or points. This can be better understood by analyzing the maps $\Gamma_{l\alpha \alpha} ( \kvec  )$ presented in Fig.~\ref{fig:NNN_GammaK}. 

For low chemical potentials, the $| \Gamma_{l\alpha \alpha}  ( \kvec ) |$ is more concentrated around the $\Gamma$ point of the Brillouin zone [Fig.~\ref{fig:NNN_GammaK}(a)], which is a clear consequence of long-range pairing, associated with the length scale determined by $\vec{\delta}_{nnn}$. In this case, the Fermi surface is primarily concentrated around the maximum of the order parameter $| \Gamma_{l\alpha \alpha}  ( \kvec )|$  and the superconducting gap is significantly enhanced. For intermediate values of $\mu$, the magnitude of $| \Gamma_{l\alpha \alpha} ( \mathbf{k} )|$ is visibly reduced [Fig.~\ref{fig:NNN_GammaK}(b), see the scale on the right], and the Fermi surface barely overlaps with the regions of maximal order parameter. As a consequence, a dip appears in the order parameter presented in Fig.~\ref{fig:NNN}. Eventually, for chemical potentials near the van Hove singularity — where the reentrance of superconductivity is observed — the Fermi surface overlaps with the high-amplitude regions of the order parameter near the edges of the first Brillouin zone [Fig.~\ref{fig:NNN_GammaK}(c)]. This behavior is well captured by the effective SC gap $\tilde{\Delta} \left( \mathbf{k} \right)$ shown in Fig.~\ref{fig:NNN_Dos_Gap}(a).
\begin{figure}[!htp]
    \centering
    \includegraphics[width=0.49\linewidth]{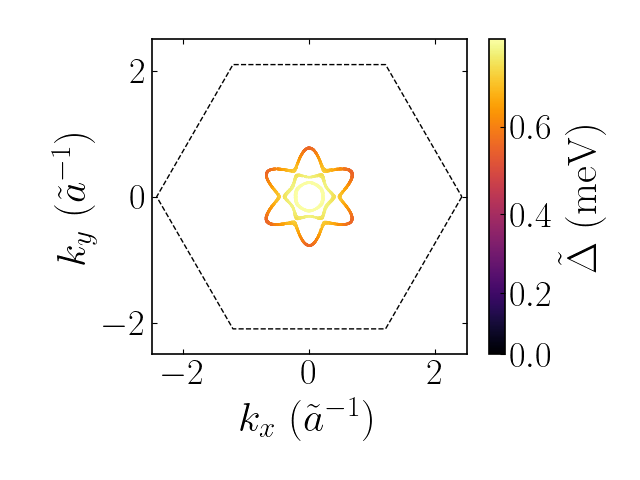}
    \put(-107,85){(a)}
    \includegraphics[width=0.49\linewidth]{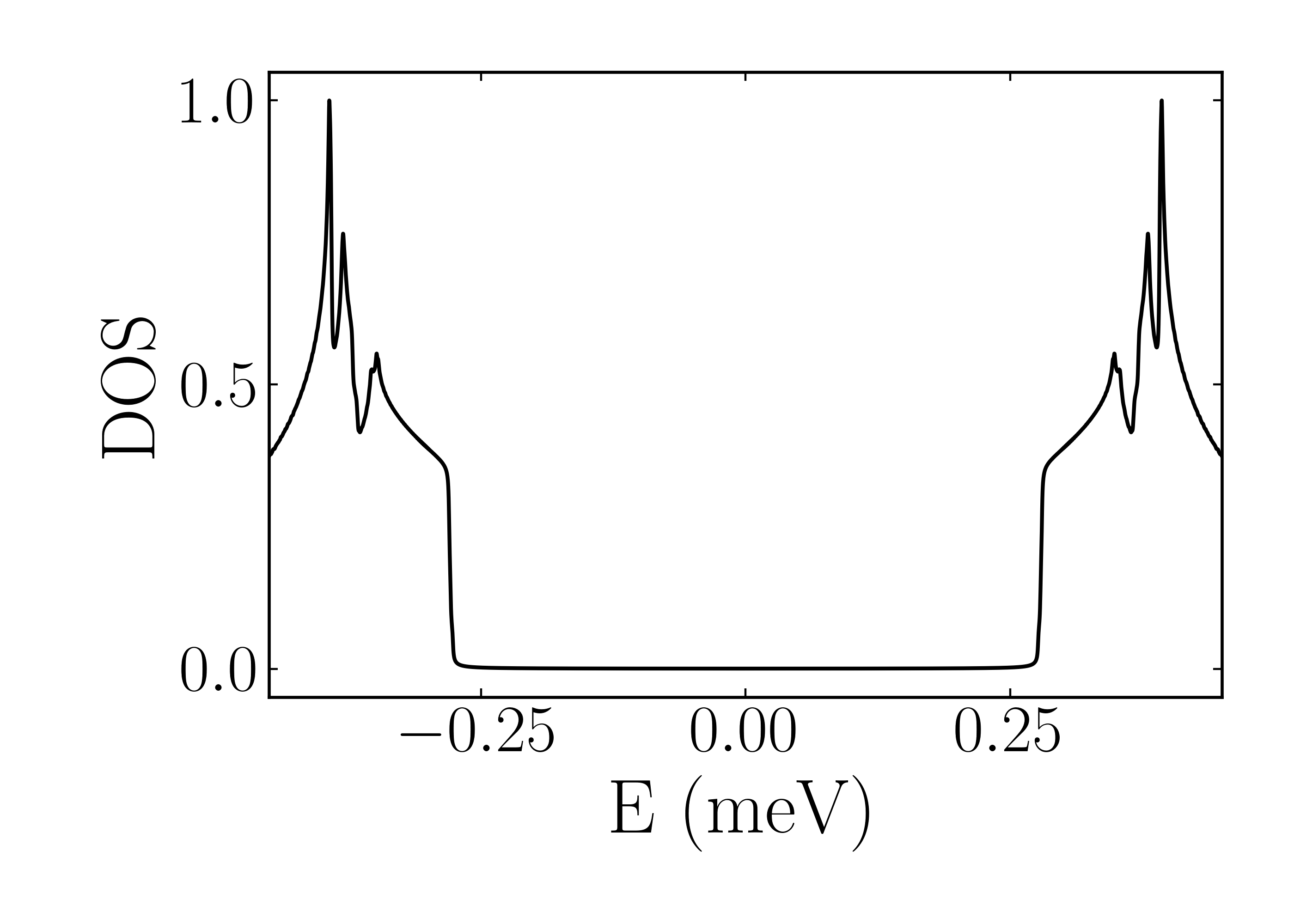}
    \put(-107,85){(b)}

    \includegraphics[width=0.49\linewidth]{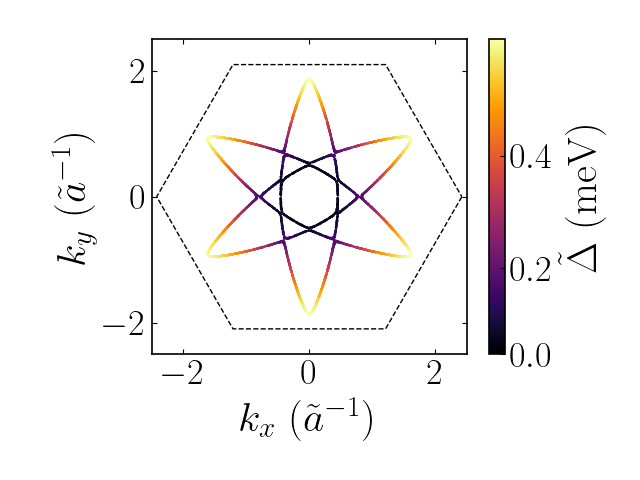}
    \put(-107,85){(c)}
    \includegraphics[width=0.49\linewidth]{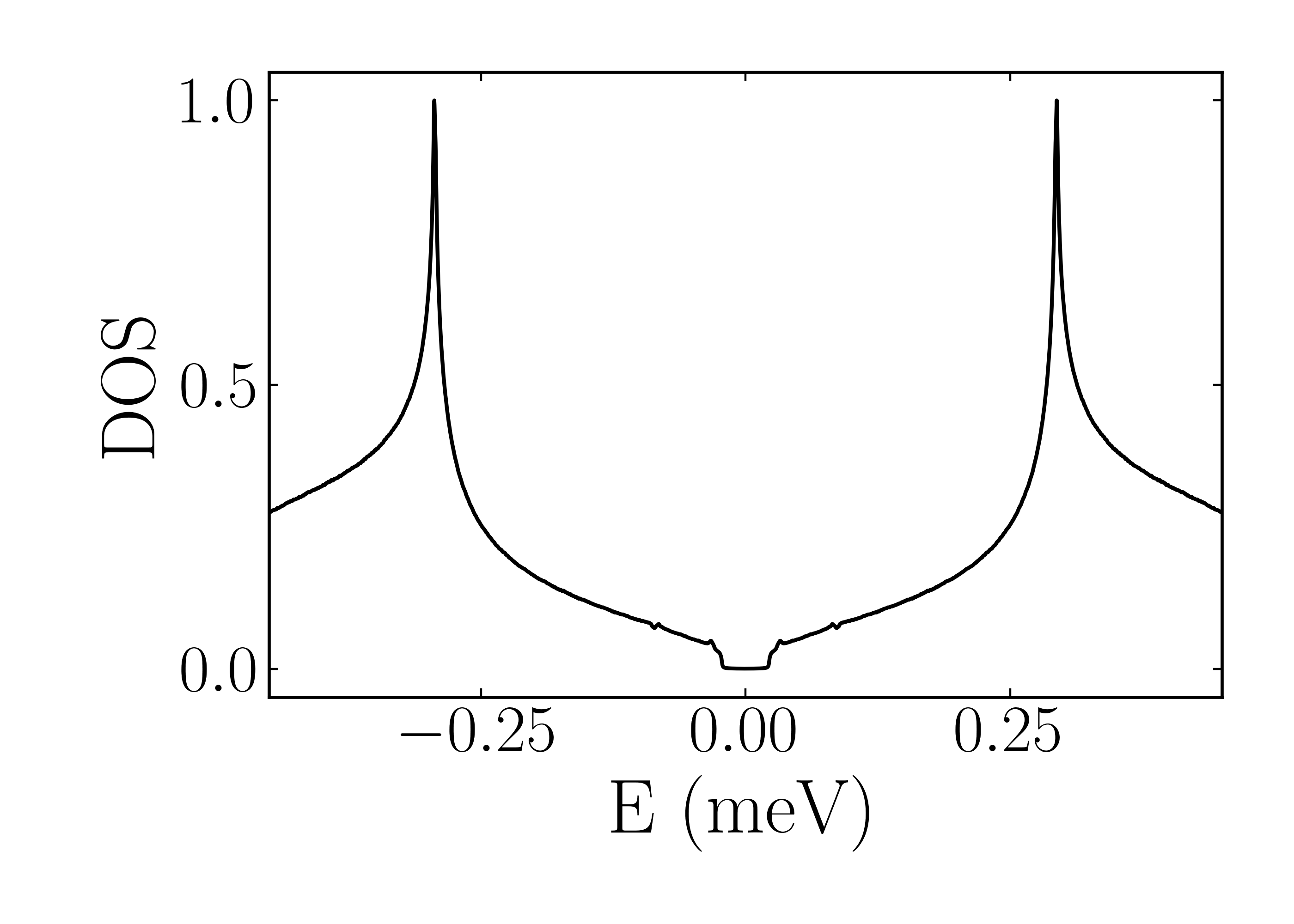}
    \put(-107,85){(d)}
    \caption{(a,c) Effective superconducting gap at the Fermi surface and (b,d) the corresponding DOS. Results for $J = 50$~meV, $J_{nnn} = 75$~meV and (a,b) $\mu = 31$~meV and (c,d) $\mu = 85$~meV. }
    \label{fig:mix_Dos_Gap}
\end{figure}
In this chemical potential (electron concentration) regime, it is particularly interesting to note that the order parameter for each orbital undergoes a sign change, clearly visible in the upper row of Fig.~\ref{fig:NNN_GammaK}(c), and is accompanied by a vanishing amplitude in the order parameter (lower row). This behavior leads to a nodal lobe [Fig.~\ref{fig:NNN_Dos_Gap}(a)], which in turn results in a $V$-shaped DOS dependence presented in Fig.~\ref{fig:NNN_Dos_Gap}(b). We point out that the appearance of the nodal lobe is a direct consequence of long-range coupling mechanism and we attribute it to a nodal $s$-wave pairing.
Finally, we should emphasize that the symmetry of the superconducting gap remains unchanged under group-theoretical classification. The order parameter continues to transform according to the $A_1$ IR, maintaining invariance under all operations of the $C_{6v}$ point group. Note, however, that this apparent geometric simplicity does not eliminate the possibility of superconductivity characterized by nodal lines/points.

\begin{figure*}
    \centering
    \includegraphics[width=0.4\linewidth]{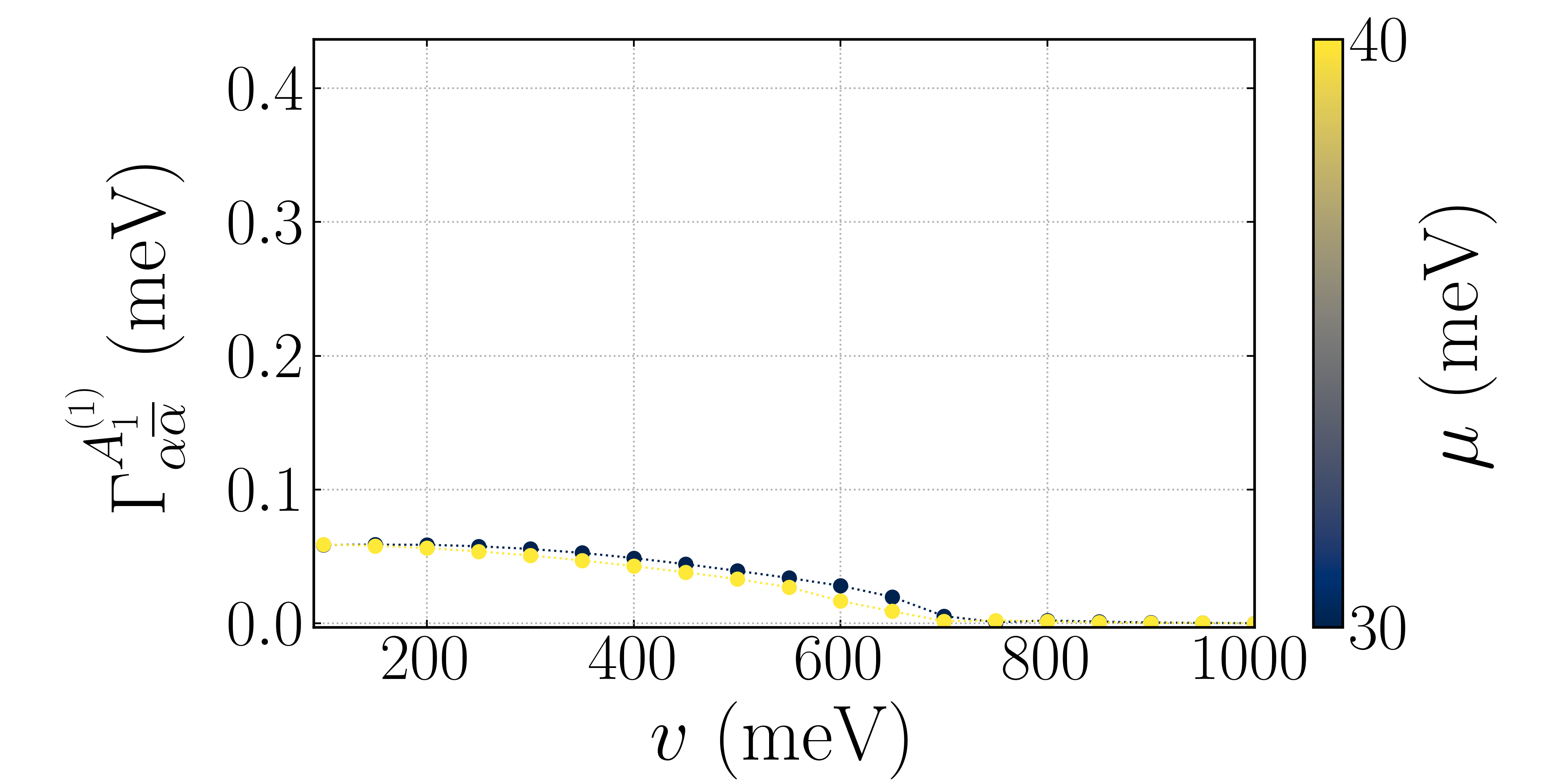}
    \put(-180,100){(a)}
    \includegraphics[width=0.4\linewidth]{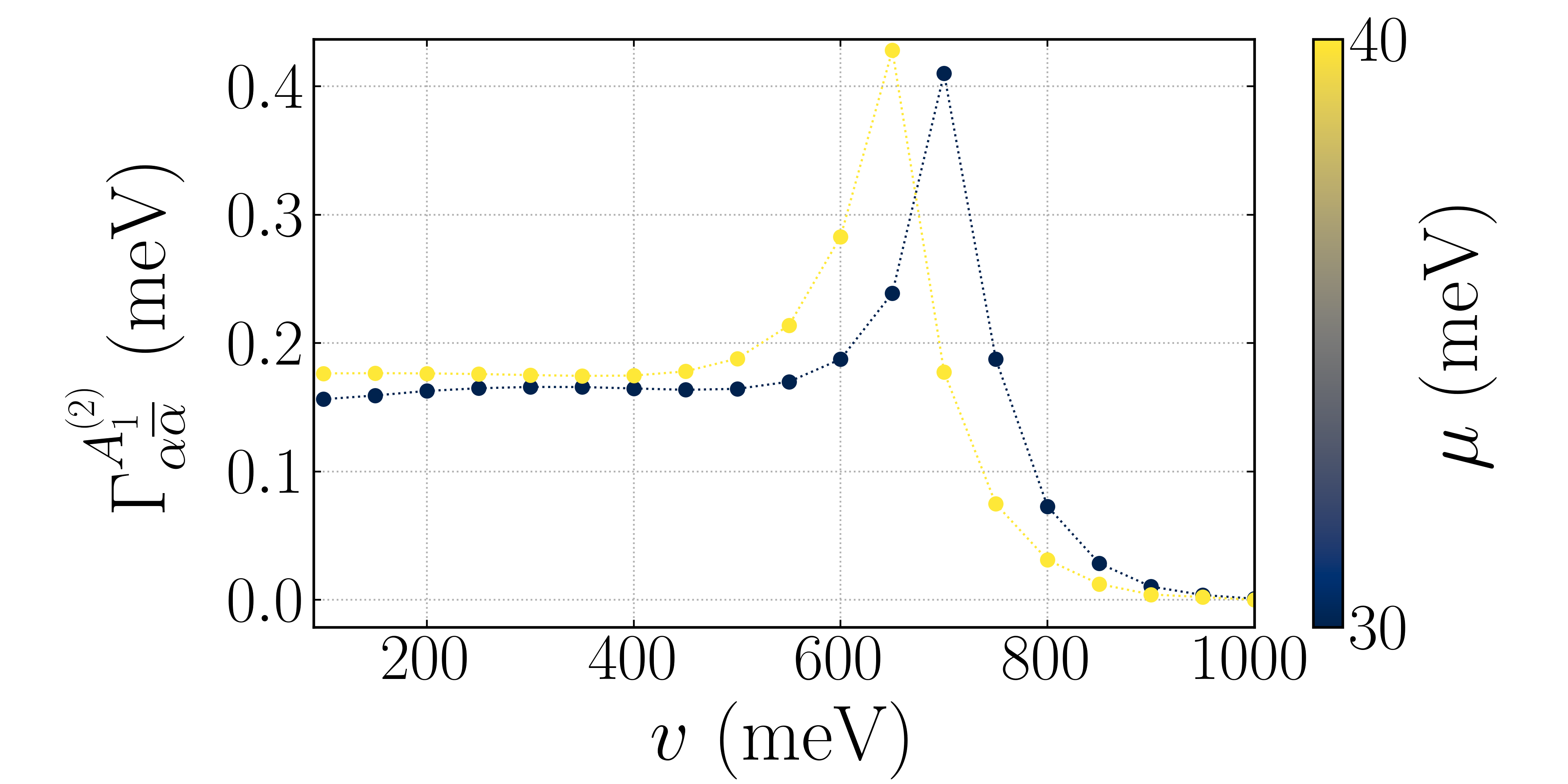}
     \put(-180,100){(b)}

    \includegraphics[width=0.4\linewidth]{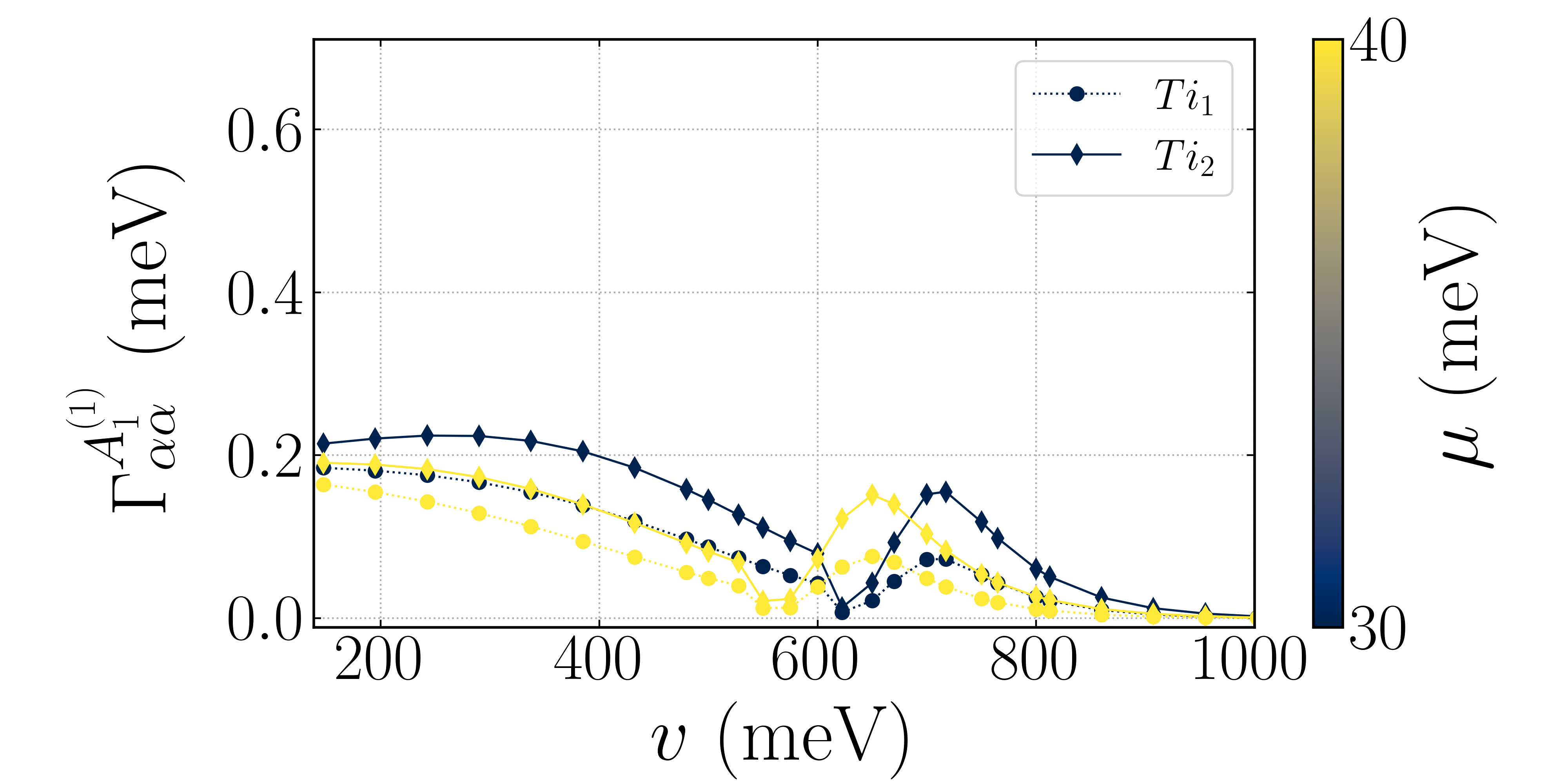}
    \put(-180,100){(c)}
    \includegraphics[width=0.4\linewidth]{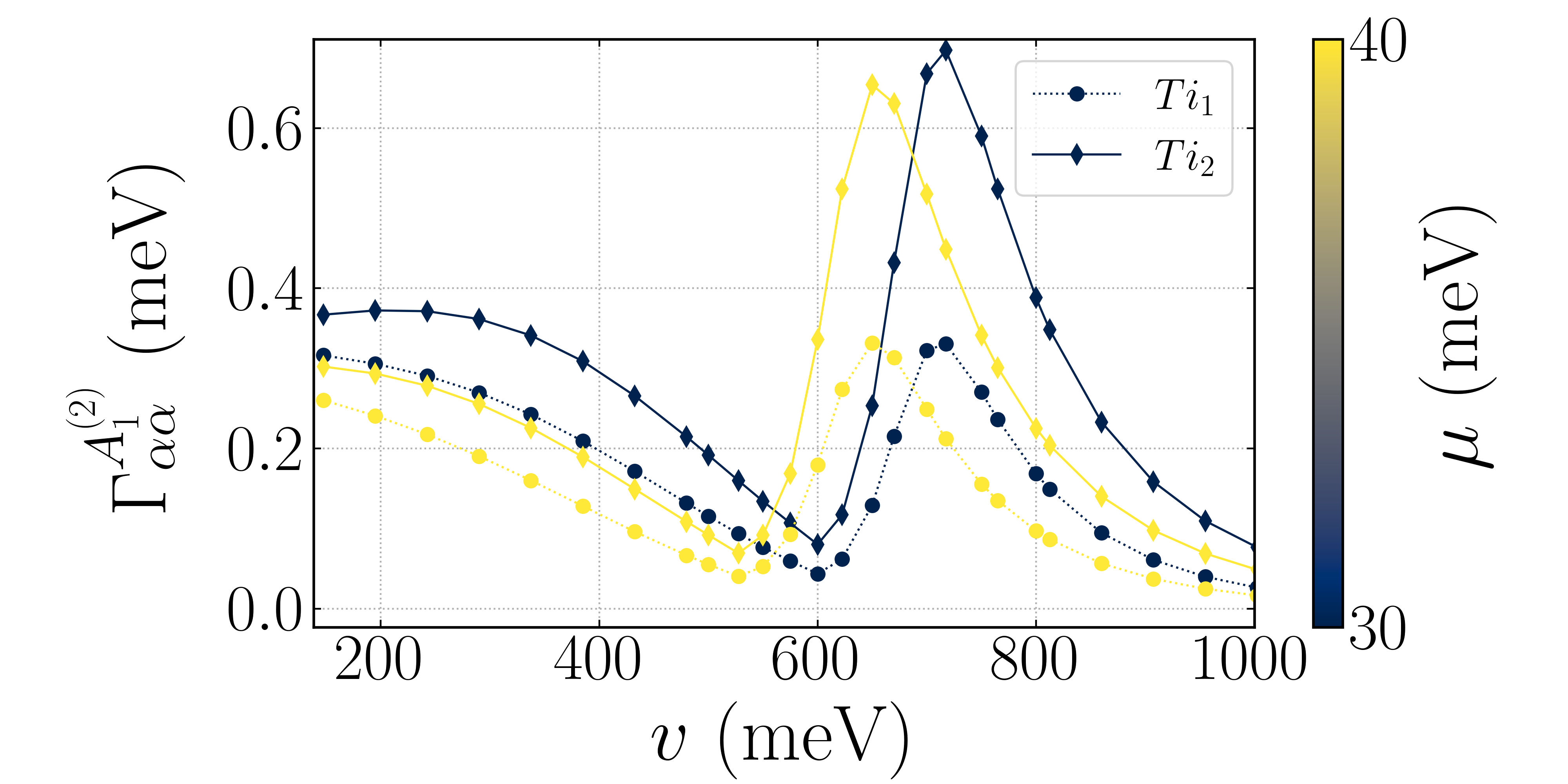}
    \put(-180,100){(d)}
    \caption{Dependence of the symmetry-resolved coupling parameters as a function of potential difference between $Ti$ layers for (a,b) solely interlayer pairing $J = 150$~meV, $J_{nnn} = 0$ and (c,d) solely intralayer coupling $J = 0$, $J_{nnn} = 150$~meV. }
    \label{fig:electricFieldDome}
\end{figure*}

\subsection{Mixed pairing}
\label{subsec:Mix}

After analyzing the impact of the interlayer and intralayer coupling channels individually on the superconducting phase diagram, we now study their interplay. For this purpose, we set finite values for both \( J \) and \( J_{nnn} \) and consider three cases: \( J > J_{nnn} \), \( J = J_{nnn} \), and \( J < J_{nnn} \), which should provide a general overview of the evolution of the superconducting state as \( J \) and \( J_{nnn} \) vary. It should be noted that if \( J \neq J_{nnn} \), we effectively model an anisotropic pairing mechanism, since the in-plane (intralayer) pairing energies differ from the out-of-plane (interlayer) ones. As proposed in Ref.~\onlinecite{BealMond_anistoropic_1996}, such anisotropy could lead to an enhancement of the critical temperature compared to the isotropic case. In the calculations, we set \( J_{nnn} = 75~\text{meV} \) and vary the energy \( J \). The ratio between intra- and interorbital Cooper pair hopping is maintained at a fixed value, as in previous sections, i.e., \( J' = 0.1J \) and \( J'_{nnn} = 0.1J_{nnn} \). Consistently, we set the Hubbard interaction energies \( U = V = 0 \).

In Fig.~\ref{fig:mix} we present symmetry-resolved superconducting order parameter in both inter- and intralayer pairing occurring simultaneously. We see that, the ratio \( J / J_{nnn} \) does not change the qualitative behavior of the system within the presented range of parameters. Instead, it only tunes the magnitudes in \textit{both} pairing channels, indicating a strong coupling between the two. We also observe that the superconducting phase diagram changes in the presence of mixed pairing. In particular, the interlayer contributions $\Gamma_{l\alpha \bar{\alpha}}$ now exhibit a noticeable dome in the low-concentration regime, in contrast to the interlayer-only coupling scenario. Moreover, we note that the intralayer contribution from \( \Gamma_{\alpha \alpha}^{A_1^{(1)}} \) is visibly smoothed and no longer vanishes in the medium-concentration regime. This leads to the disappearance of the reentrance feature previously reported for superconductivity mediated by the intralayer channel.

For completeness, in Fig.~\ref{fig:mix_Dos_Gap} we present the effective gaps and density of states (DOS) for two different chemical potentials. Notably, throughout the entire chemical potential range, the superconductivity remains fully gapped. We attribute this to the impact of the interlayer coupling channel, which is not affected by a nodal structure, as shown in Fig.~\ref{fig:NN_GammaK}. Interestingly, we observe that the DOS exhibits multiple coherence peaks for \( \mu = 31~\text{meV} \) (Fig.~\ref{fig:mix_Dos_Gap}(b)) and a substantial smoothing around the coherence peaks for \( \mu = 85~\text{meV} \) (Fig.~\ref{fig:mix_Dos_Gap}(d)).

\begin{figure*}[!htb]
\includegraphics[width=0.33\linewidth]{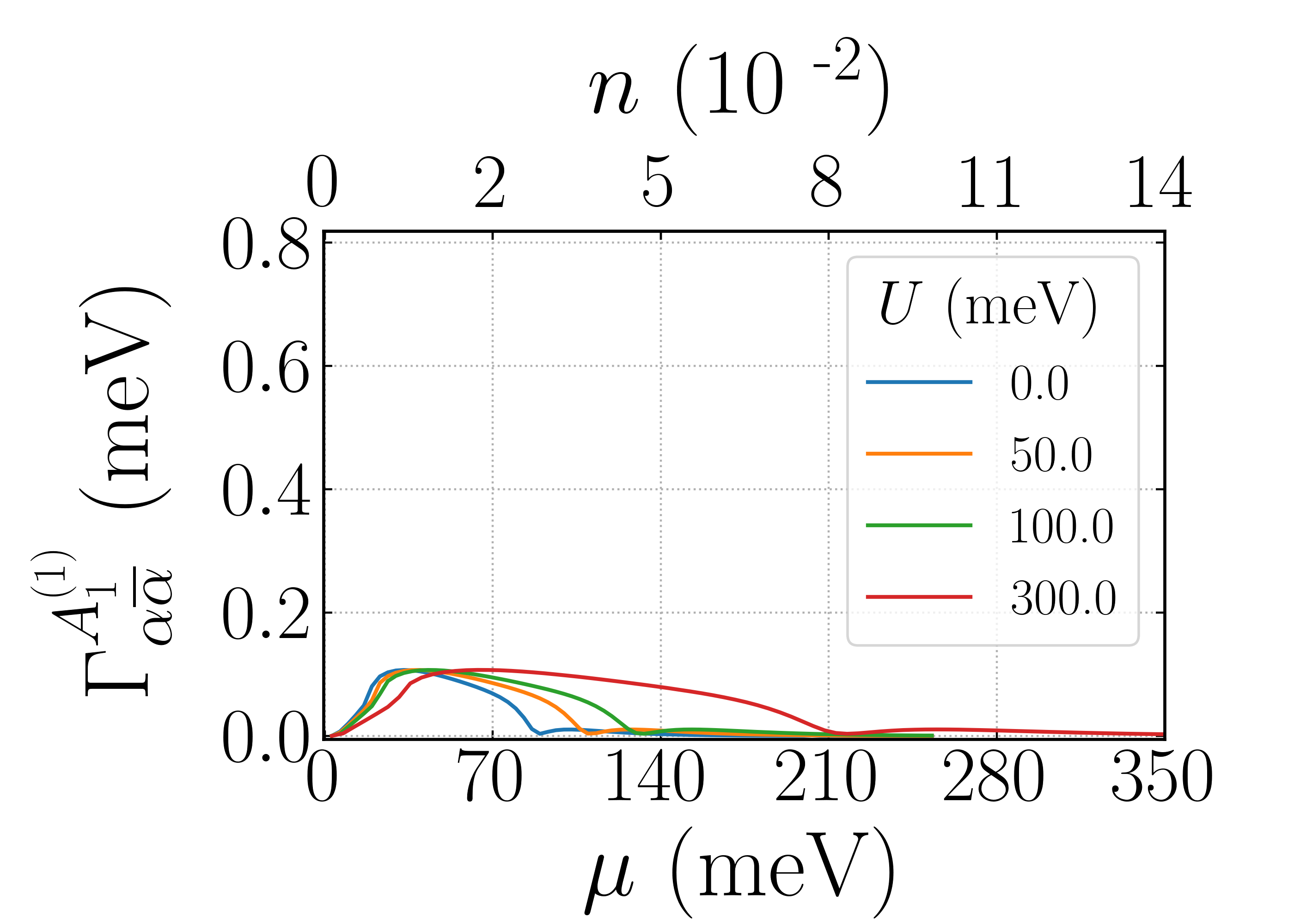}  \put(-145,100){(a)}
\includegraphics[width=0.33\linewidth]{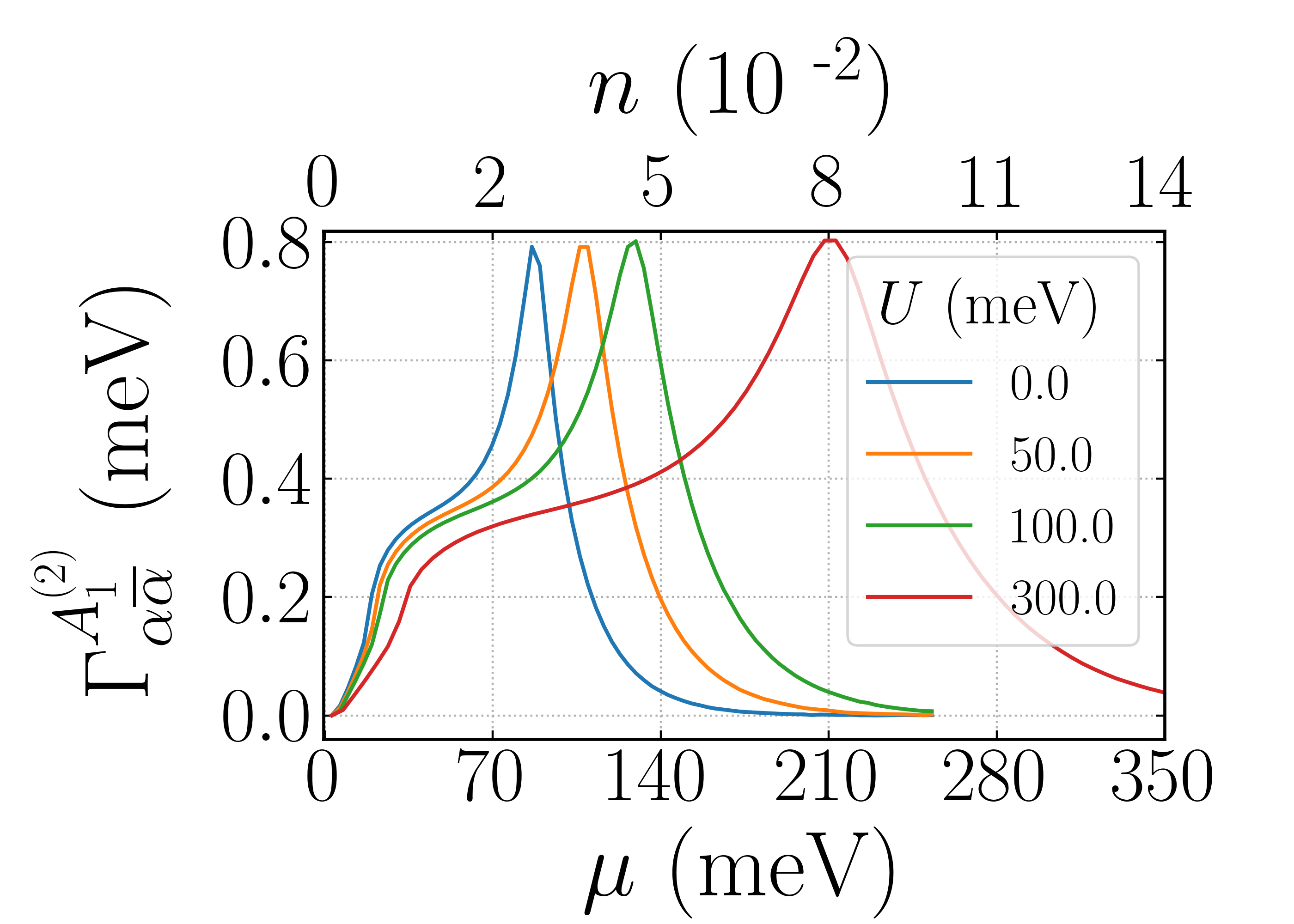}  \put(-145,100){(b)}
\includegraphics[width=0.33\linewidth]{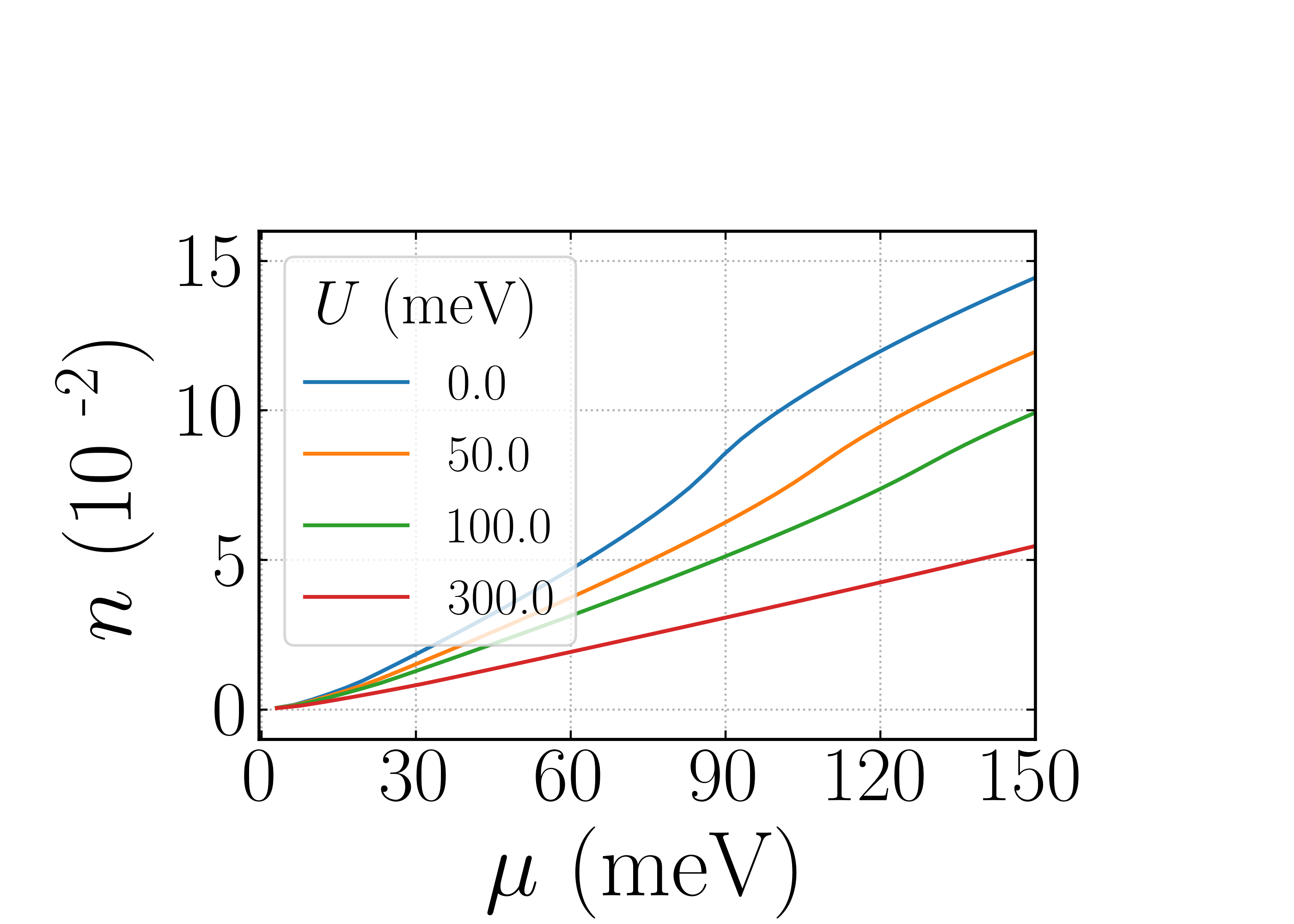}
\put(-145,100){(c)}
\caption{(a,b) Magnitudes of symmetry-resolved superconducting pairing amplitudes and (c) electronic concentration as a function of chemical potential $\mu$. For presented results, Hubbard interaction only changes the energy scale and does not impose changes on the superconducting phase diagram. Results presented for $J = 170$~meV.}
\label{fig:DomeAndConcentrationHub}
\end{figure*}

\subsection{Electric field}
\label{subsec:ElectricField}

Since exceptional features in LAO/STO are believed to stem from its spin-orbit coupling arising from the breaking of inversion symmetry at the interface, we now study the impact of an electric field perpendicular to the interface — see Eq.~\ref{eq:H_externalElectric}. We performed calculations for two distinct scenarios: (i) solely interlayer coupling with \( J = 150~\text{meV} \), and (ii) solely intralayer coupling with \( J_{nnn} = 150~\text{meV} \). We vary \( v \), the potential difference between layers, in the range \([100, 1000]~\text{meV}\) for two chemical potentials, \(\mu = 30\) and \(\mu = 40~\text{meV}\), corresponding to low carrier concentrations. The results for both scenarios are presented in Fig.~\ref{fig:electricFieldDome}(a,b) and (c,d), respectively.

Surprisingly, for the interlayer pairing [Fig.~\ref{fig:electricFieldDome} (a,b)] we observe a nonmonotonic behavior with a profound peak around $v = 600$~meV. Intuitively, one might anticipate that increasing the energetic separation between the two layers would reduce the pairing between the layers, thereby causing a monotonic suppression of superconducting properties. However, before the pairing becomes negligible, the system experiences a Lifshitz transition associated with the complete filling of the band corresponding to the energetically lower-lying Ti layer. This transition induces a van Hove singularity in the density of states, which enhances the superconducting pairing strength. In particular, we observe a similar behavior for both projections in Fig.~\ref{fig:NN}, where the \( A_1^{(1)} \) projection is suppressed near the singularity [Fig.~\ref{fig:NN}(a)], whereas the \( A_1^{(2)} \) projection is markedly enhanced [Fig.~\ref{fig:NN} (b)].

The behavior of the intralayer pairing channel exhibits a pronounced sensitivity to the potential difference $v$. As $v$ increases, the pairing projections in both layers decrease in magnitude. This behavior is seen up to $v \approx 600$~meV, despite a redistribution of carriers that enhances the electron density in the Ti$_2$ layer at the expense of Ti$_1$. This counterintuitive trend is similar to the dependence of the superconducting order parameter on the chemical potential in the vicinity of van Hove singularity, presented in Fig.~\ref{fig:NNN}. Similarly, above $v \approx 600$~meV, we observe a reentrance of the superconducting properties, which fade out for high electric fields. We conclude that tuning the Rashba spin-orbit coupling by applying an external electric field is limited by the suppression of the order parameter in strong electric fields.

\subsection{Hubbard interaction}
\label{subsec:HubbardInteraction}
To finalize our analysis, we present the dependence of superconducting properties on the on-site repulsion energy $U$ and $V$ in the Hubbard model assuming $U = V$. The results are shown in Fig.~\ref{fig:DomeAndConcentrationHub} where only the projection $A_1$ in the interlayer pairing scenario is presented. The corresponding ratios of both projections, $A_1^{(1)}$ and $A_1^{(2)}$, remain constant across all values of $\mu$, leading us to conclude that onsite repulsion primarily acts as a scaling factor for the chemical potential within the parameter range considered in this study. This is clearly illustrated in Fig.~\ref{fig:DomeAndConcentrationHub}(c), which shows the carrier concentration as a function of the chemical potential for various values of $U$. The nonlinearity observed in the $n(\mu)$, particularly around the inflection point, is attributed to the presence of a van Hove singularity, reflecting a reduction in the number of available states within the lowest-lying Kramers pair.

\section{Summary}
\label{sec:Summary}
In this paper, we performed self-consistent calculations of the superconducting order parameter for the 2DEG at the LAO/STO interface oriented along the (111) crystallographic direction.
Our theoretical model incorporates the kinetic term, spin-orbit coupling, interfacial symmetry breaking, and trigonal lattice strain, all included within the tight binding approximation.  Electronic interactions, including real-space attractive pairing and on-site Hubbard repulsion, are treated within the mean-field approximation. To characterize the superconducting phase, we performed a rigorous group-theoretical analysis of the order parameter, providing insight into the symmetry of the superconducting state. The dependence of the order parameters on the chemical potential (or electron concentration) is analyzed based on maps of their distribution in wave vector space and the topology of the Fermi surface.

In the paper, we analyzed the symmetry of the superconducting order parameter in scenarios where superconductivity is driven by three mechanisms: (i) interlayer pairing, (ii) intralayer pairing, and (iii) a combination of both interlayer and intralayer pairing. In all cases, the order parameter transforms according to the fully symmetric $A_1$ irreducible representation of the $C_{6v}$ point group.

Within the interlayer pairing scenario, we found that the superconducting phase is characterized by a fully gapped quasiparticle spectrum exhibiting extended $s$-wave symmetry and an enhancement of the gap near the van Hove singularity. On the other hand, when intralayer coupling is included, the order parameter exhibits a pronounced dome-shaped dependence on the chemical potential. The similarity to the experimental measurements indicates that intralayer pairing between next-nearest neighbors may play a crucial role, supporting the hypothesis of a fundamentally two-dimensional character of superconductivity at the (111) LAO/STO interface. Interestingly, in the case of purely intralayer coupling, we showed that the superconducting state is characterized by the nodal lines in the high carrier concentration regime leading to the characteristic $V$-shaped of DOS. We hypothesize that this $V$-shaped behavior, triggered by the nodal lobe, could be observed for even lower chemical potentials if the pairing interaction range is longer. The estimated critical temperatures lie within the experimentally measured ranges, demonstrating that the model can quantitatively reproduce experimental observations by appropriate tuning of the real-space pairing strengths $J$ and $J_{nnn}$.

Finally, we explored the effect of symmetry breaking at the interface determined by the external electric field applied perpendicular to it. Remarkably, the order parameter magnitude varies nonmonotonically with increasing electric field strength— a behavior we link to a van Hove singularity arising from the complete filling of one of the Ti layers. We demonstrated that the superconducting gap driven by interlayer and intralayer channels differs qualitatively under the electric field. Specifically, intralayer coupling exhibits both suppression and reentrance of superconductivity as a function of field strength, whereas interlayer coupling maintains a finite gap over a broad range of field values before vanishing. We propose that this distinct behavior can serve as experimental probes to identify the dominant pairing channel at the (111) LAO/STO interface.

\begin{acknowledgments}
This research was partially supported by a subsidy from the Polish
Ministry of Science and Higher Education and the program „Excellence initiative – research university” for the AGH University. Computing infrastructure PLGrid (HPC Centers: ACK Cyfronet AGH) was used within computational grant no. PLG/2024/017213.
 
\end{acknowledgments}

\appendix
\section{Renormalization of hopping parameters}
\label{sec:HoppingRenormalization}
An electric field perpendicular to the interface, as a consequence of symmetry breaking, not only induces a potential difference between the Ti layers but also modifies the hopping terms. This renormalization is given by $h_{\kvec l l'}^{\alpha \alpha'}\left( v \right)$ in Eq. \eqref{eq:H_externalElectric} and reads
\begin{equation}
\footnotesize
\begin{aligned}
    &h_{\kvec ll'}^{\alpha \overline{\alpha}} (v)= \eta_p \frac{V_{p d \pi}(\sqrt{2})^{7 / 4}}{\sqrt{15}} \times \\
    &\begin{pmatrix}
    0 & -2 i e^{i \frac{3}{2} k_y} \sin \left(\frac{\sqrt{3}}{2} k_x\right) & 1-e^{\frac{i}{2}\left(\sqrt{3} k_x+3 k_y\right)} \\
    2 i e^{i \frac{3}{2} k_y} \sin \left(\frac{\sqrt{3}}{2} k_x\right) & 0 & 1-e^{-\frac{i}{2}\left(\sqrt{3} k_x-3 k_y\right)} \\
    -1+e^{\frac{i}{2}\left(\sqrt{3} k_x+3 k_y\right)} & -1+e^{-\frac{i}{2}\left(\sqrt{3} k_x-3 k_Y\right)} & 0
    \end{pmatrix}
\end{aligned}
\label{eq:H_electricNearest}
\end{equation}
for hopping between nearest neighbors. \\
For the next nearest neighbors, it takes the form
\begin{equation}
h_{\kvec ll'}^{\alpha \alpha} (v) = h_{\kvec ll'}^{\pi} (v) + h_{\kvec ll'}^{\sigma} (v),
\end{equation}
where

\begin{widetext}
\begin{subequations}
\footnotesize
\begin{align}
    &h_{\kvec ll'}^{\pi} (v) = \eta_p \frac{2 i}{\sqrt{15}} V_{p d \pi} \begin{pmatrix}
    0 & -\left(\sin \left(\kappa_1\right)+\sin \left(\kappa_2\right) + 2 \sin \left(\kappa_3 \right)\right) & \left(\sin \left(\kappa_1\right)+ 2 \sin \left(\kappa_2 \right) + \sin \left(\kappa_3\right)\right) \\
    \left(\sin \left(\kappa_1\right)+\sin \left(\kappa_2\right) + 2 \sin \left(\kappa_3 \right)\right) & 0 & - \left( 2 \sin \left(\kappa_1 \right) \sin \left(\kappa_2\right) + \sin \left(\kappa_3\right)\right) \\
    -\left(\sin \left(\kappa_1\right)+ 2 \sin \left(\kappa_2 \right) + \sin \left(\kappa_3\right)\right) & \left(2 \sin \left(\kappa_1 \right) + \sin \left(\kappa_2\right)+\sin \left(\kappa_3\right)\right) & 0    
    \end{pmatrix}, & \\
    &h_{\kvec ll'}^{\sigma} (v) = \eta_p \frac{2 i}{\sqrt{15}} \sqrt{3} V_{p d \sigma} \begin{pmatrix}
    0 & \left(\sin \left(\kappa_1\right)+\sin \left(\kappa_2\right)\right) & -\left(\sin \left(\kappa_1\right)+\sin \left(\kappa_3\right)\right) \\
    -\left(\sin \left(\kappa_1\right)+\sin \left(\kappa_2\right)\right) & 0 & \left(\sin \left(\kappa_2\right)+\sin \left(\kappa_3\right)\right) \\
    \left(\sin \left(\kappa_1\right)+\sin \left(\kappa_3\right)\right) & -\left(\sin \left(\kappa_2\right)+\sin \left(\kappa_3\right)\right) & 0    
    \end{pmatrix}.
\end{align}
\label{eq:H_externalNextNearest}
\end{subequations}
\end{widetext}
and we define
\begin{eqnarray}
    \kappa_1 &=& -\frac{\sqrt{3}}{2}k_x + \frac{3}{2}k_y, \\
    \kappa_2 &=& -\frac{\sqrt{3}}{2}k_x - \frac{3}{2}k_y, \\
    \kappa_3 &=& \sqrt{3}k_x. 
\end{eqnarray}

In Eqs. \eqref{eq:H_externalNextNearest}, $V_{p d \pi}$ and $V_{p d \sigma}$ are Slater-Koster integrals, associated with $\pi$ and $\sigma$ overlap between orbitals respectively. For LAO/STO they take values $V_{pd\pi} = 28$~meV and $V_{pd\sigma} = -65$~meV. The dependence of the parameter $\eta_p$ on the electric field is given by
\begin{equation}
    \eta_p \simeq \frac{\sqrt{3}v}{a_0} \frac{1}{10~eV \cdot nm^{-1}},
\end{equation}
where $a_0 = 3.905$~nm is the lattice constant and $v$ is the potential difference between Ti layers. Full derivation can be found in Supplementary Material of Ref. \cite{Trama_BerryCurvature_2022}.


%

\end{document}